\documentclass{aa}  
\usepackage{ae,aecompl}
\usepackage{adjustbox}
\usepackage{graphicx}
\usepackage{txfonts}
\usepackage{hyperref}
\usepackage{xcolor}

\hypersetup{colorlinks,linkcolor={blue},citecolor={blue},urlcolor={blue}}  
\usepackage{graphicx}
\usepackage{epsf,palatino,url,wrapfig,array,setspace,amsmath,amssymb,fancyhdr,multirow,lscape,appendix,rotating}
\usepackage{graphics,epsfig,txfonts}
\usepackage{threeparttable}
 
\newcommand{\hsp}{3HSP~J095507.9+355101}
\newcommand{\nustar}{\emph{NuSTAR}}

\newcommand{\fermi}{\emph{Fermi}}

\newcommand{\code}{{\tt LeHaMoC}}

\begin{document}

   \title{LeHaMoC\thanks{Instructions for downloading the code, accessing online documentation, and reproducing applications presented in this paper can be found at \url{https://github.com/mariapetro/LeHaMoC} Github repository.}: A versatile time-dependent lepto-hadronic modeling code for high-energy astrophysical sources}
   \titlerunning{LeHaMoC}
   \author{S. I. Stathopoulos \inst{1},
          M. Petropoulou  \inst{1,2},
          G. Vasilopoulos \inst{1,2},          
          A. Mastichiadis  \inst{1}
          }       
   \institute{Department of Physics, National and Kapodistrian University of Athens, University Campus Zografos, GR 15784, Athens, Greece, \email{stamstath@phys.uoa.gr}
 \and
    Institute of Accelerating Systems \& Applications, University Campus Zografos, Athens, Greece}
   \date{Received: 24 June 2023; accepted: 25 December 2023}

 
  \abstract
   {Recent associations of high-energy neutrinos with active galactic nuclei (AGN) have revived the interest in leptohadronic models of radiation from astrophysical sources. The rapid increase in the amount of acquired multi-messenger data will require fast numerical models that may be applied to large source samples.}
   {We develop a time-dependent leptohadronic code, \code , that offers several notable benefits compared to other existing codes, such as versatility and speed.}
   {\code \ solves the Fokker-Planck equations of photons and relativistic particles (i.e. electrons, positrons, protons, and neutrinos) produced in a homogeneous magnetized source that may also be expanding. The code utilizes a fully implicit difference scheme that allows fast computation of steady-state and dynamically evolving physical problems.}
   {We first present test cases where we compare the numerical results obtained with \code \ against exact analytical solutions and numerical results computed with ATHE$\nu$A, a well-tested code of similar philosophy but a different numerical implementation. We find a good agreement (within 10-30\%) with the numerical results obtained with ATHE$\nu$A without evidence of systematic differences. We then demonstrate the capabilities of the code through illustrative examples. First, we fit the spectral energy distribution from a jetted AGN in the context of a synchrotron-self Compton model and a proton-synchrotron model using Bayesian inference. Second, we compute the high-energy neutrino signal and the electromagnetic cascade induced by hadronic interactions in the corona of NGC 1068.}
   {\code \ is easily customized to model a variety of high-energy astrophysical sources and has the potential to become a widely utilized tool in multi-messenger astrophysics.}


   \maketitle
%
\section{Introduction}
A wide variety of astrophysical sources, including supernova remnants (SNRs), microquasars, active galactic nuclei (AGN), and gamma-ray bursts (GRBs), display non-thermal emission spectral energy distributions (SEDs). This emission, which depending on the type of the source, may span from radio wavelengths to $\gamma$-ray energies, requires the presence of relativistic particles with extended energy distributions. Modeling of the broadband emission has been a widely used method for probing indirectly the physical conditions of these astrophysical accelerators, such as particle density, magnetic field strength, bulk motion speed, and others.

Since the discovery of $\gamma$-ray emission from AGN (see the reviews \cite{2016CRPhy..17..594D,doi:10.1146/annurev-astro-081913-040044} and references therein), the production mechanism of such energetic photons has been a matter of debate. Two schools of thought have been developed over the years while trying to explain the emission properties of $\gamma$-ray emitting sources. The first one postulates that the broadband emission can be fully attributed to energetic electrons accelerated in the source, with some contribution from secondary electron-positron pairs produced through photon-photon pair production (leptonic models). According to the second school, accelerated protons in the source could also make a significant contribution to the high-energy emission (leptohadronic models), either directly via proton synchrotron radiation \citep{Aharonian2000, Muecke2001}, or indirectly via the radiation of secondary pairs produced in proton-photon interactions \citep{Mannheim_1993}, or via neutral pion decay \citep{2005ApJ...630..186R}. High-energy neutrinos, produced via the pion decay chain, are a natural outcome of this model class. 

The recent discovery by the IceCube Collaboration of high-energy neutrino emission (at a significance of $\sim 4.2\sigma$) associated with the prototype Seyfert II galaxy NGC 1068 \citep{2022Sci...378..538I} presents a new set of challenges for understanding the underlying physics of these sources. Notably, the observed GeV $\gamma$-ray emission from NGC 1068 is relatively low compared to the neutrino luminosity \citep{2022Sci...378..538I}, which raises intriguing questions about the mechanisms responsible for this disparity and the potential production sites \citep[see e.g.,][for different explanations]{Murase_2020, 2022arXiv220702097I, 2022ApJ...939...43E}. The other astronomical source that has been associated with a significance larger than $\sim 3 \sigma$ with IceCube neutrinos is the blazar\footnote{Blazars are a subclass of AGN with relativistic jets viewed at a small angle \citep{Urry1995}.} TXS~0506+056 \citep{IceCube2018_flare, IceCube2018_archival}. These observations, in addition to the discovery of the diffuse astrophysical neutrino flux by IceCube \citep{icecube2013}, have revived the interest in leptohadronic source models.

Leptohadronic radiative models are intrinsically more complex than leptonic ones. The main radiative processes in the latter model class are synchrotron radiation, inverse Compton scattering, photon-photon pair-production, and relativistic bremsstrahlung (which is relevant for sources with dense ionized gas). In leptohadronic models, one has to take also into account the interactions of relativistic protons with low-energy photons and/or gas. Both processes lead to the production of pions that decay into photons, pairs, and neutrinos. Photons and pairs can then initiate electromagnetic cascades in the source via photon-photon pair production, thus producing more targets for proton-photon interactions. The complexity of leptohadronic models lies in (i) the non-linear coupling of the various particle populations (through the so-called kinetic equations i.e., partial differential equations with respect to time and energy), (ii) the wide range of timescales probed by the physical processes, and (iii) the numerical modeling of the secondary particle production spectra. The adoption of different approaches for the modeling of secondary particle injection may yield 10-30\% differences in the numerical results obtained with different radiative transfer codes (\citet{2022icrc.confE.979C}, Cerutti et al., in prep.).

Various radiative transfer codes have been employed to study the complex emission processes and interpret the diverse observational properties of non-thermal emitting sources, including recent neutrino observations~\citep[e.g.][]{DMPR12, B13, LeHa-Paris, AM3, SOPRANO}. The rapid increase in multi-messenger observations and modeling of larger and more diverse source samples requires high computation efficiency and code adaptability. In this work, we present \code \, (Lepto-Hadronic Modeling Code),  a novel radiative transfer code for non-thermal emitting sources. While \code\, is inspired by the modeling code ATHE$\nu$A \citep{mastichiadis1995synchrotron, DMPR12}, it offers several new key features: significantly shorter code execution times for steady-state problems due to an implicit finite difference scheme \citep{chang1970a}, modeling of expanding sources with varying physical conditions, the inclusion of proton-proton inelastic collisions, and fitting capabilities. Our primary goal is to provide a versatile and efficient numerical code to the community.

This paper is structured as follows. In Sec.~\ref{sec:code} we outline the range of processes incorporated into \code \,  and describe the methodology employed to solve the kinetic equations for each particle species in the source. In Sec.~\ref{sec:performance} we provide an assessment of the code's performance and accuracy. In Sec.~\ref{sec:tests} we present indicative cases that are compared against analytical results and numerical results obtained with the well-tested code ATHE$\nu$A \citep{DMPR12}.  In Sec.~\ref{sec:applications} we showcase two astrophysical applications of \code. Firstly, we fit the SED of a blazar that has been proposed as a potential high-energy neutrino emitter using a Bayesian inference approach. Secondly, we employ \code \, to compute the high-energy neutrino signal and derive the electromagnetic cascade resulting from hadronic interactions within the corona of NGC 1068. By examining these specific cases, we demonstrate the versatility and utility of \code \, in describing different astrophysical sources. Finally, we present the conclusions of this work in Sec.~\ref{sec:conclusions}.

\section{Code description}\label{sec:code}
We consider a scenario in which relativistic charged particles (electrons and/or protons) are injected into a spherical blob that contains a magnetic field and may also adiabatically expand. Charged particles are subject to various physical processes that lead to the production of photons and secondary particles, such as relativistic electron-positron pairs, neutrons, and neutrinos. The processes we consider are
\begin{itemize}
  \item Electron and positron synchrotron radiation ($e,syn$),
  \item Proton synchrotron radiation ($p,syn$),
  \item Electron and positron inverse Compton scattering ($ics$),
  \item Synchrotron self-absorption ($ssa$),
  \item Photon-photon pair creation ($\gamma\gamma$),
  \item Proton-photon pion production ($ p\gamma,\pi$),
  \item Proton-photon (Bethe-Heitler) pair production  ($bh$),
  \item Proton-proton collisions ($pp$),
  \item Adiabatic losses ($ad$). 
\end{itemize}

All particle species are assumed to be isotropically distributed inside the source and to be fully described by the relevant distribution function $N_i$ at time $t$, which is a function of the charged particle  Lorentz factor $\gamma_i$, or frequency $\nu$ for photons and neutrinos. We trace the evolution of charged particles and photons inside the source as a function of energy and time by numerically solving a system of coupled integrodifferential (kinetic) equations that are summarized below.

\begin{itemize}
  \item Electrons and positrons,
\begin{equation}
\frac{\partial N_{e} }{\partial t}+\frac{N_{e}}{\tau_{e,esc}}+\mathcal{L}_{e}^{syn}+\mathcal{L}_{e}^{ics}+\mathcal{L}_{e}^{ad} =Q_{e}^{inj}+Q_{e}^{p\gamma,\pi}+Q_{e}^{bh}+Q_{e}^{pp}+Q_{e}^{\gamma \gamma}.
\label{eq:gen_kin_eq_e}
\end{equation}

  \item Protons,
\begin{equation}
\frac{\partial N_{p} }{\partial t}+\frac{N_{p}}{\tau_{p,esc}}+\mathcal{L}_{p}^{syn}+\mathcal{L}_{p}^{ad}+\mathcal{L}_{p}^{bh}+\mathcal{L}_{p}^{p\gamma,\pi}+\mathcal{L}_{p}^{pp} =Q_{p}^{inj}.
\label{eq:gen_kin_eq_p}
\end{equation}

  \item Photons,
\begin{equation}
\frac{\partial N_{\gamma} }{\partial t}+\frac{N_{\gamma}}{\tau_{\gamma,esc}}+\mathcal{L}_{\gamma}^{ssa}+\mathcal{L}_{\gamma}^{\gamma\gamma}+\mathcal{L}_{\gamma}^{ad} =Q_{\gamma}^{e,syn}+Q_{\gamma}^{p,syn}+Q_{\gamma}^{ics}+Q_{\gamma}^{p\gamma,\pi}+Q_{\gamma}^{pp}.
\label{eq:gen_kin_eq_nu}
\end{equation}

  \item Neutrinos,
\begin{equation}
\frac{\partial N_{\nu} }{\partial t}+\frac{N_{\nu}}{\tau_{\nu,esc}} = Q_{\nu}^{p\gamma, \pi}+Q_{\nu}^{pp}.
\label{eq:gen_kin_eq_g}
\end{equation}
\end{itemize}
where we account for the contribution of each flavor separately.

In the equations above $Q^{j}_{i}$ expresses the production rate of particle species $i$ due to the process $j$, while the terms designated with the subscript $inj$ indicate the injection rate of accelerated particles (primaries) into the source. Similarly, the $\mathcal{L}^j_i$ term represents the net loss rate of species $i$ due to the physical process $j$. We have also introduced $\tau_{i,esc}$ which is the physical escape timescale of species $i$ from the source. In most cases, this timescale is taken to be a multiple of the source's light-crossing time $t_{cr}=R(t)/c$, with $R(t)$ being the radius of the spherical blob. Alternatively, if the escape probability is related to the gyroradius of the particle in the source, $\tau_{i,esc}$ may depend on the particle energy. We allow for both a time-dependent and energy-dependent escape timescale in our code.

We neglect the neutron component and assume that the decay of pions is instantaneous, so we do not treat them using kinetic equations. Detailed expressions of the operators used in the kinetic equations can be found in Appendix~\ref{appA}.

To solve the kinetic equations presented in the previous sections we use the fully implicit difference scheme proposed by \citet{chang1970a}. The stiff differential equations that arise in the context of astrophysical jets, where the various timescales associated with the physical processes can range by many orders of magnitude, are particularly well-suited for this numerical scheme. 
The specific implicit numerical scheme involves the discretization of the kinetic equations in time and the Lorentz factor and frequency space. To discretize the kinetic equations in time, we use a time step that is related to the characteristic time scale of the system, which depends on the physical scenario studied. The use of an implicit scheme ensures that the solution at any time step is stable, which is useful since the time scales of different physical processes can vary widely. The discretization in energy is achieved by using a logarithmic energy grid that allows us to accurately capture the behavior of the distribution over a wide range of energies. The energy grid size is a parameter that can be defined by the user. Each discretized equation forms a tridiagonal matrix which we solve using Thomas algorithm \citep{thomas1949elliptic}. 

To illustrate how the code works, we demonstrate the discretization into time $t_i$ and Lorentz factor $\gamma_j$ of the kinetic equation for the electron species (Eq.~\ref{eq:gen_kin_eq_e}),

\begin{equation}
V_{1,j}N_{e,j-1}^{i+1}+V_{2,j}N_{e,j}^{i+1}+V_{3,j}N_{e,j+1}^{i+1}=N_{e,j}^i,
\label{eq:k_e_N_e_n_s}
\end{equation}
where the coefficients are
\begin{equation}
V_{1,j} = 0,
\label{eq:V1_coef_N_e_n_s}
\end{equation}
\begin{equation}
V_{2,j} = 1+\frac{\Delta t}{\tau_{e,esc}}+\frac{\Delta t}{\Delta \gamma_j}\sum_{p}\left(\frac{d\gamma}{dt}\right)_{p, j}^{i+1},
\label{eq:V2_coef_N_e_n_s}
\end{equation}
\begin{equation}
V_{3,j} = -\frac{\Delta t}{\Delta \gamma_j}\sum_{p}\left(\frac{d\gamma}{dt}\right)_{p, j+1}^{i+1},
\label{eq:V3_coef_N_e_n_s}
\end{equation}
where $(d\gamma/dt)_p$ are the energy losses to an electron due to the $p$ process.

\section{Code performance and testing} 
\subsection{Code performance}\label{sec:performance} 

We demonstrate the performance and accuracy of the numerical code through three steady-state leptonic runs. In all cases, electrons are injected in the source with a power-law distribution, $N_e(\gamma_e)\propto\gamma_e^{-s_e}$, for $\gamma_{e, \max} \ge \gamma \ge \gamma_{e, \min}$. For the adopted parameters,
electrons, which are continuously injected with $\gamma_e \gg 1$, are completely cooled down to $\gamma_e \gtrsim 1$ (i.e. fast cooling regime) via synchrotron radiation (SYN) or inverse Compton (IC) scattering on a fixed blackbody radiation field. A synchrotron self-Compton scenario (SSC) where electrons are still cooled due to synchrotron losses is also considered. In all tests, we examine the relationship between the Lorentz factor grid resolution with the code execution time and the ratio of the bolometric photon luminosity $L^{bol}_\gamma$, and the relativistic electron luminosity at injection, $L^{inj}_{e}$. In the fast cooling regime, all energy injected into relativistic electrons should be radiated away, hence the two luminosities should be the same. The parameters used for the performance test are displayed in Table \ref{table0}. It is important to note that our code is not parallelized and utilizes a single CPU core for computations. All the tests were performed in 12th Gen Intel® Core™ i5-1235U $\times$ 12 processor.

\renewcommand{\arraystretch}{1.2}
\begin{table}[]
    \centering
\caption{Parameter values for the code performance tests in Sec.~\ref{sec:performance}.}
    \begin{threeparttable}
    \begin{tabular}{@{}lccc@{}}
\hline
Parameters & \multicolumn{3}{c}{Tests} \\
\hline  
           & SYN & IC & SSC \\
\hline 
$R_0$ [cm]  & \multicolumn{3}{c}{$10^{15}$}  \\
$B_0$ [G]  & $10^3$ & - & $10^3$ \\
$T_{\rm BB}$ [K] &  - & $10^5$ &  - \\
$\gamma_{e,\min}$    & \multicolumn{3}{c}{1}  \\
$\gamma_{e,\max}$   & \multicolumn{3}{c}{$10^9$}  \\
$s_e$ & \multicolumn{3}{c}{2}   \\
$L^{inj}_e~[\rm erg \ s^{-1}]$  & \multicolumn{3}{c}{$3.1\cdot10^{40}$}  \\
\hline
    \end{tabular}  
    \label{table0}
    \end{threeparttable}
\end{table}

\begin{figure}
\centering
\includegraphics[width=0.45\textwidth]{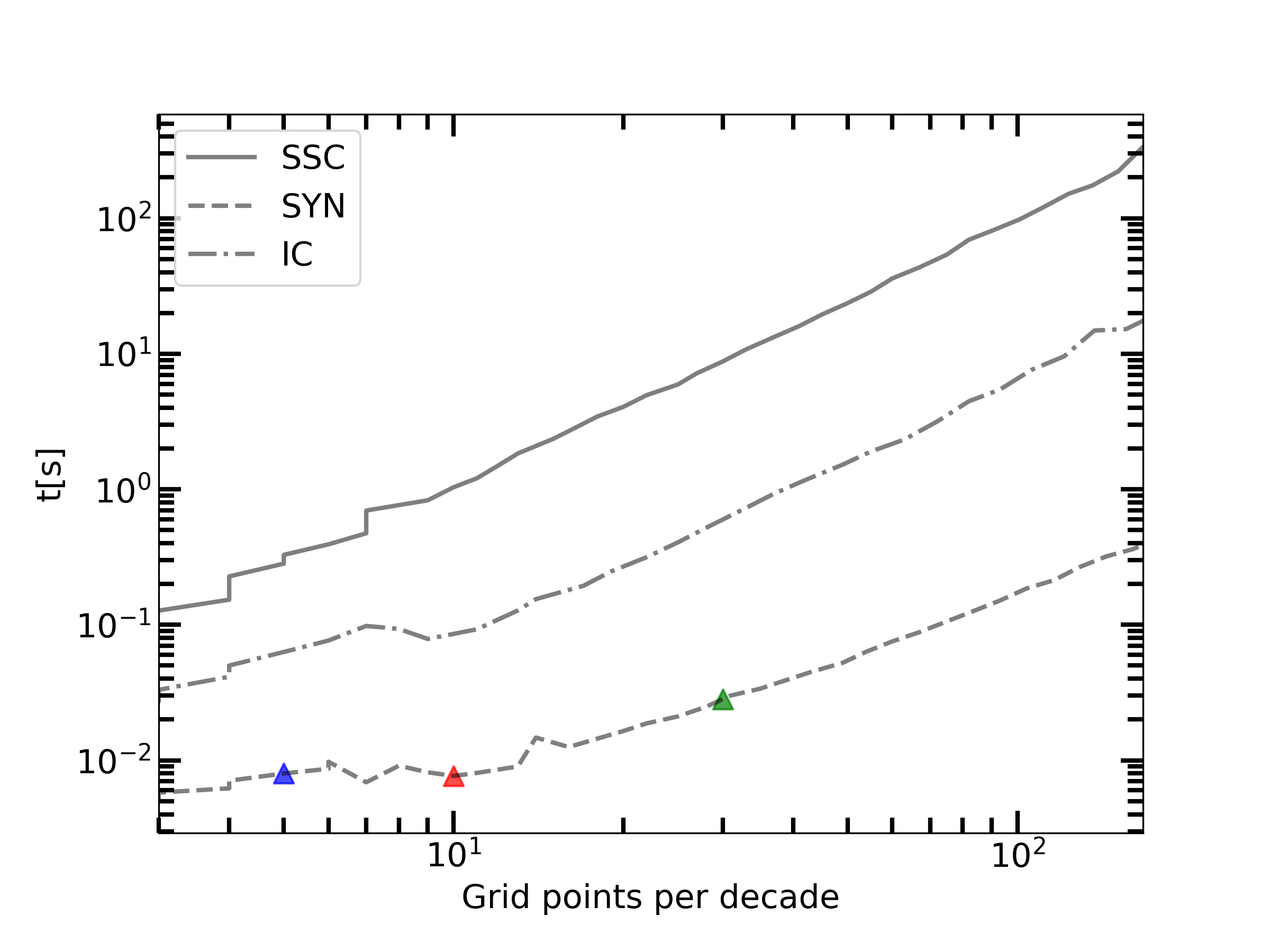}\\
\includegraphics[width=0.45\textwidth]{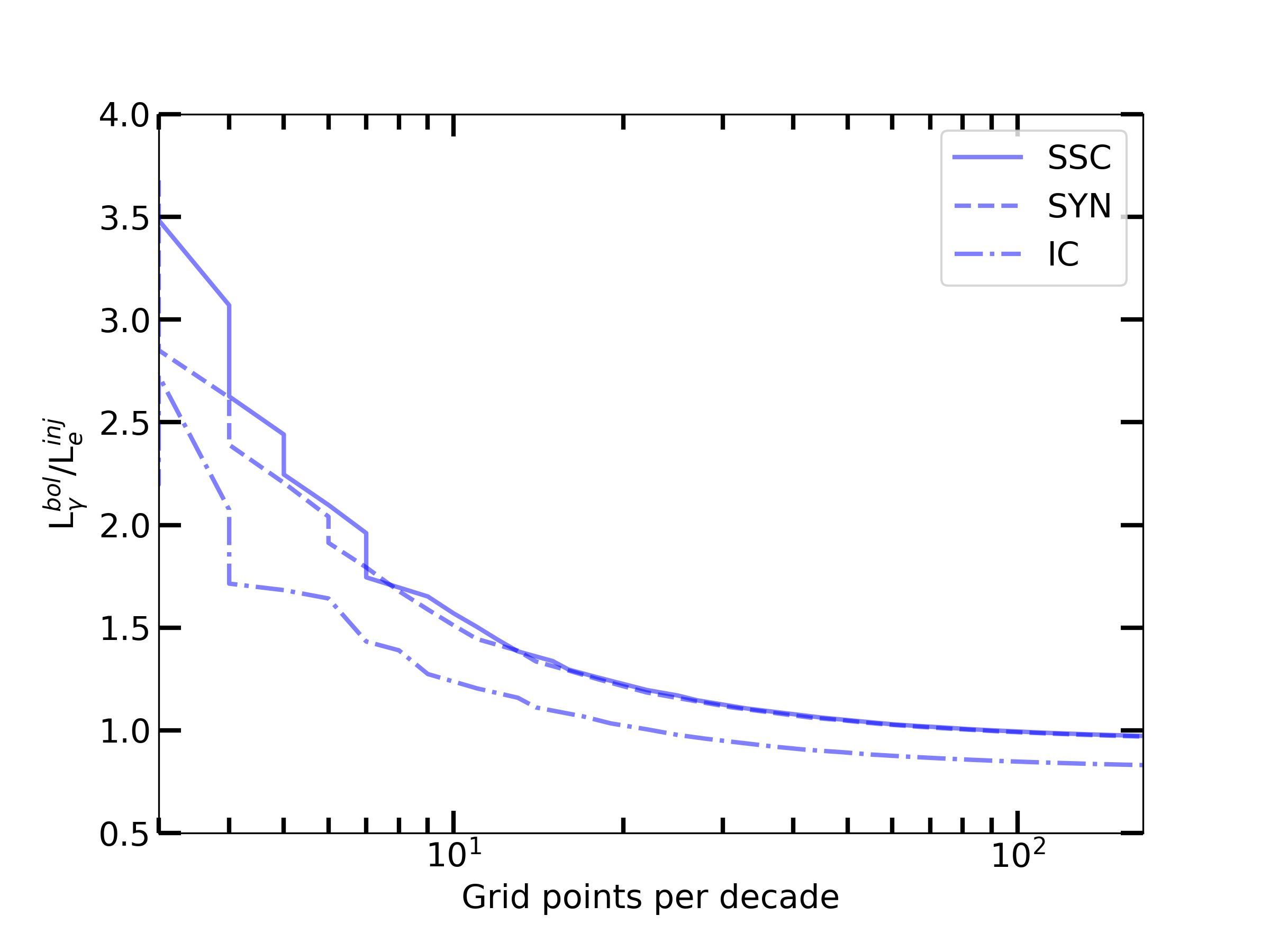}\\
\caption{ Impact of grid resolution on code execution time and luminosity balance. Top panel: Relation between the code execution time and the number of grid points per logarithmic decade in particle energy for three scenarios of fast cooling electrons described in the text (see Sec.~\ref{sec:performance}). Colored triangles indicate three choices for the number of grid points per decade that are used for the results shown in Fig.~\ref{fig:N_el_vs_grid_points}.
Bottom panel: Ratio of bolometric photon luminosity to electron injection luminosity versus the number of grid points per logarithmic decade. A ratio of one indicates the expected energy balance between electrons and photons.}
\label{fig:grid_points}       
\end{figure}

The number of grid points ($n_{\rm ppd}$) directly affects the execution time for all three scenarios. The synchrotron scenario is less impacted because the synchrotron emissivity (see Eq.~\ref{eq:Syn_emis}) is a less complicated expression than the inverse Compton scattering emissivity, which involves an integral on the target photon field (see Eq.~\ref{eq:IC_emis}), which may also evolve with time. The top panel of Fig.~\ref{fig:grid_points} shows that increasing the number of grid points per logarithmic decade increases the time to run the code for all scenarios. This trend is generally expected as more grid points require the computation of emissivity at each point, resulting in longer execution times. However, the relationship between the number of grid points and execution time is not linear. In the complete synchrotron cooling scenario, the execution time increases more rapidly when  $n_{\rm ppd}\gtrsim 20$. In the complete inverse Compton cooling scenario, the execution time scales quadratically with the number of grid points because the emissivity involves the calculation of an integral to the target photon field. The SSC scenario follows the same scaling as the inverse Compton cooling scenario since the computation of the inverse Compton emissivity dominates the execution time. 

For a typical choice of $n_{\rm ppd}=30$, the code requires approximately 10~s to reach a steady state for synchro-Compton scenarios, using a time step $dt = R_0/c$. Shorter running times can be achieved by increasing the time step. We verified that a steady-state solution can be accurately achieved with larger time steps. For example, a time step of $3\ R_0/c$ can yield accurate steady-state solutions for an SSC model, thus helping to speed up the code execution time by a factor of $\sim3$. Similar conclusions are expected for steady-state calculations including proton synchrotron radiation. The inclusion of photo-hadronic and pp interactions may raise the computation time to 5-10 minutes for the default energy resolution ($n_{\rm ppd}=20-30$).

We also evaluated the accuracy of the newly developed code by analyzing the ratio of the bolometric photon to electron injection luminosity as a function of $n_{\rm ppd}$ (see bottom panel of Fig. \ref{fig:grid_points}). As electrons are fast cooling due to synchrotron radiation in all three cases, the two luminosities should be the same. The results show that in the synchrotron and SSC scenarios, the luminosity ratio is about 4 for $n_{\rm ppd}=3$, but then decreases gradually to unity (as it should) as the number of grid points increases. The reason for this is that the increased energy resolution allows for more accurate calculations of integrals over particle distributions.         

\begin{figure}
\centering
\includegraphics[width=0.45\textwidth]{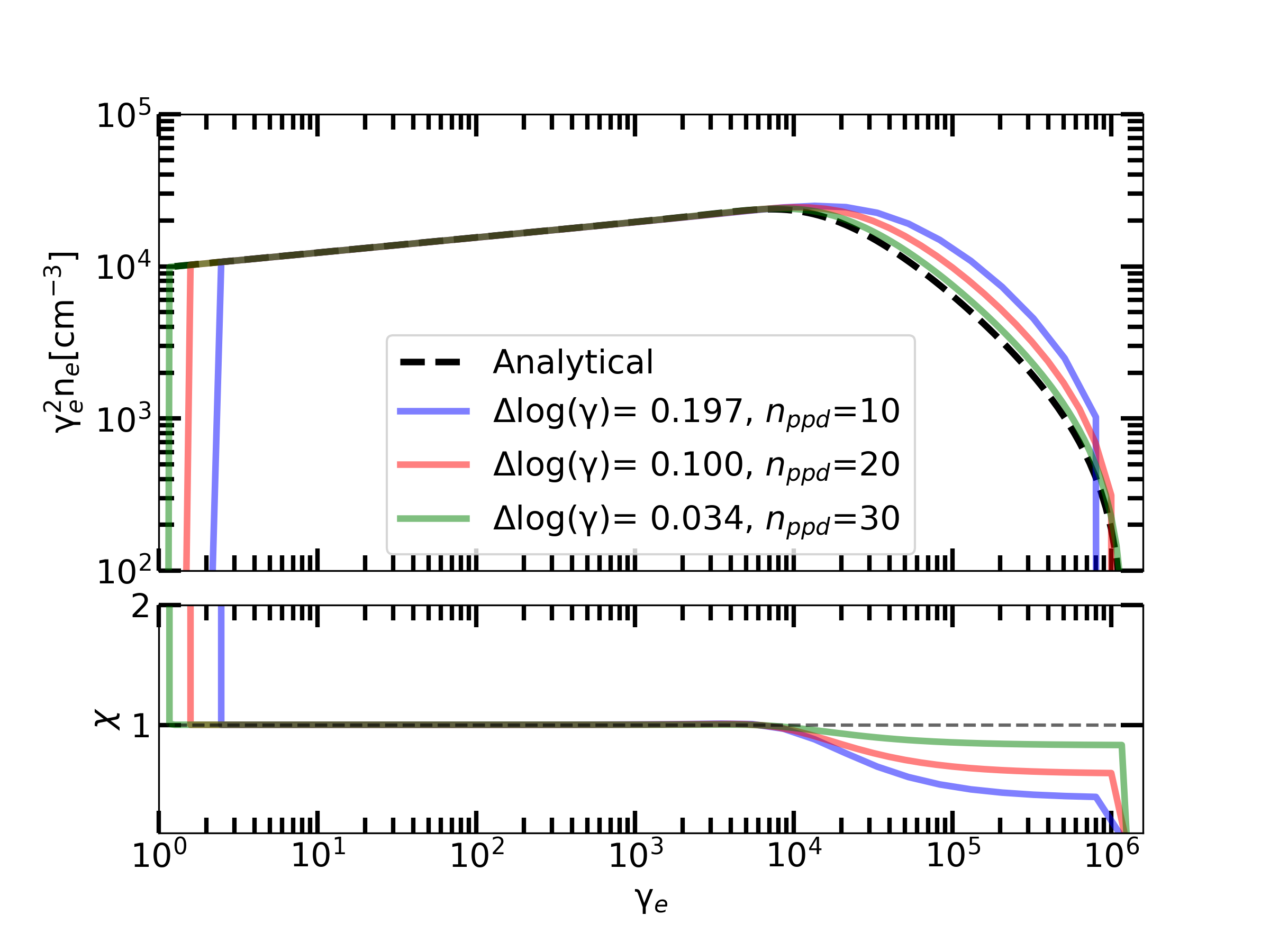}\\
\caption{Analytical and numerical comparison of a synchrotron cooling problem and the impact of grid resolution. Top panel: Comparison of numerical (solid colored lines) and analytical (dashed black line) solutions for synchrotron cooling electrons in a spherical blob with a constant magnetic field. Each numerical result was calculated using a different number of grid points per decade in the electron Lorentz factor ($n_{\rm ppd}$) as indicated in the legend (see also markers in the top panel of Fig.~\ref{fig:grid_points}). Bottom panel: Ratio $\chi$ between the numerical and the analytical solutions.}
\label{fig:N_el_vs_grid_points}       
\end{figure}

We also examine how the number of grid points affects the spectral shape of the steady-state electron distribution in a fast-cooling synchrotron scenario. For this purpose, we compare the numerical results for three choices of the energy grid resolution (marked with symbols on the top panel of Fig. \ref{fig:grid_points}) against the analytical solution of 
\cite{1996ApJ...463..555I} (see Eq.~2.26 therein). The steady-state electron distributions are displayed in Fig.~\ref{fig:N_el_vs_grid_points}, where we see that higher resolution in the energy grid leads to better agreement with the expected (analytical) solution. The adopted numerical scheme does not accurately produce the location of the cooling break if a sparse energy grid is used. According to \cite{1996ApJS..103..255P} (see their Eq.~13) the magnitude of $\Delta \gamma$ should be smaller than the typical range of variation of the kinetic equation solution at Lorentz factor $\gamma$. 

These examples illustrate the competition between accuracy and efficiency. To achieve faster computations, we have to use a coarser energy grid (typically with $n_{\rm ppd}=20-30$) and introduce a correction factor to ensure energy balance between species. Given that the synchrotron photon field is a common target field for a variety of physical processes (e.g. Compton scattering and photohadronic interactions), we determine the appropriate correction factor by running a synchrotron fast cooling case, before we perform any simulation based on the energy grid selected by the user. The spectrum of the secondaries is not affected by the number of grid points.

Finally, it is worth noting that an extremely high accuracy is not needed for most astrophysical applications since fitting uncertainties are usually the limiting factor. This will become evident later in Sec.~\ref{sec:blazar-sed} where we use the code to model the multi-wavelength spectrum of a typical jetted AGN (see also corner plots in Appendix~\ref{section:C_P}). Consequently, opting for $n_{\rm ppd}=10-20$ strikes a balance between precision and computational efficiency.

\subsection{Tests}\label{sec:tests}
We perform comparison tests against the numerical code ATHE$\nu$A \citep{mastichiadis1995synchrotron, DMPR12} in order to validate the accuracy and performance of the newly developed numerical code. In the following tests, we adopt $n_{\rm ppd}=30$, unless stated otherwise. These tests include various radiative processes that are relevant for astrophysical jets, such as synchrotron radiation, inverse Compton scattering, and proton interactions with photons. We also compare our code results with analytical calculations. In the following sections, we will provide a few examples of these comparison tests and discuss the results in detail. 

\subsubsection{Test 1: A steady-state SSC model with electron cooling}\label{Test1}
In our first comparison test, we study a synchrotron self-Compton (SSC) model with $\gamma \gamma$ absorption. The aim is to compare the photon spectrum and the pair injection rate between the ATHE$\nu$A code and \code. Electrons are injected with a power-law distribution into a spherical blob with radius $R_0$ that contains a magnetic field of strength $B_0$ and are allowed to escape on a timescale equal to the light-crossing time of the source, i.e. $t_{cr} = R_0/c$ (for the parameter values see Table~\ref{table1}).  We include synchrotron emission and self-absorption, inverse Compton scattering, and $\gamma\gamma$ pair production.

\begin{figure}
\centering
\includegraphics[width=8.5cm]{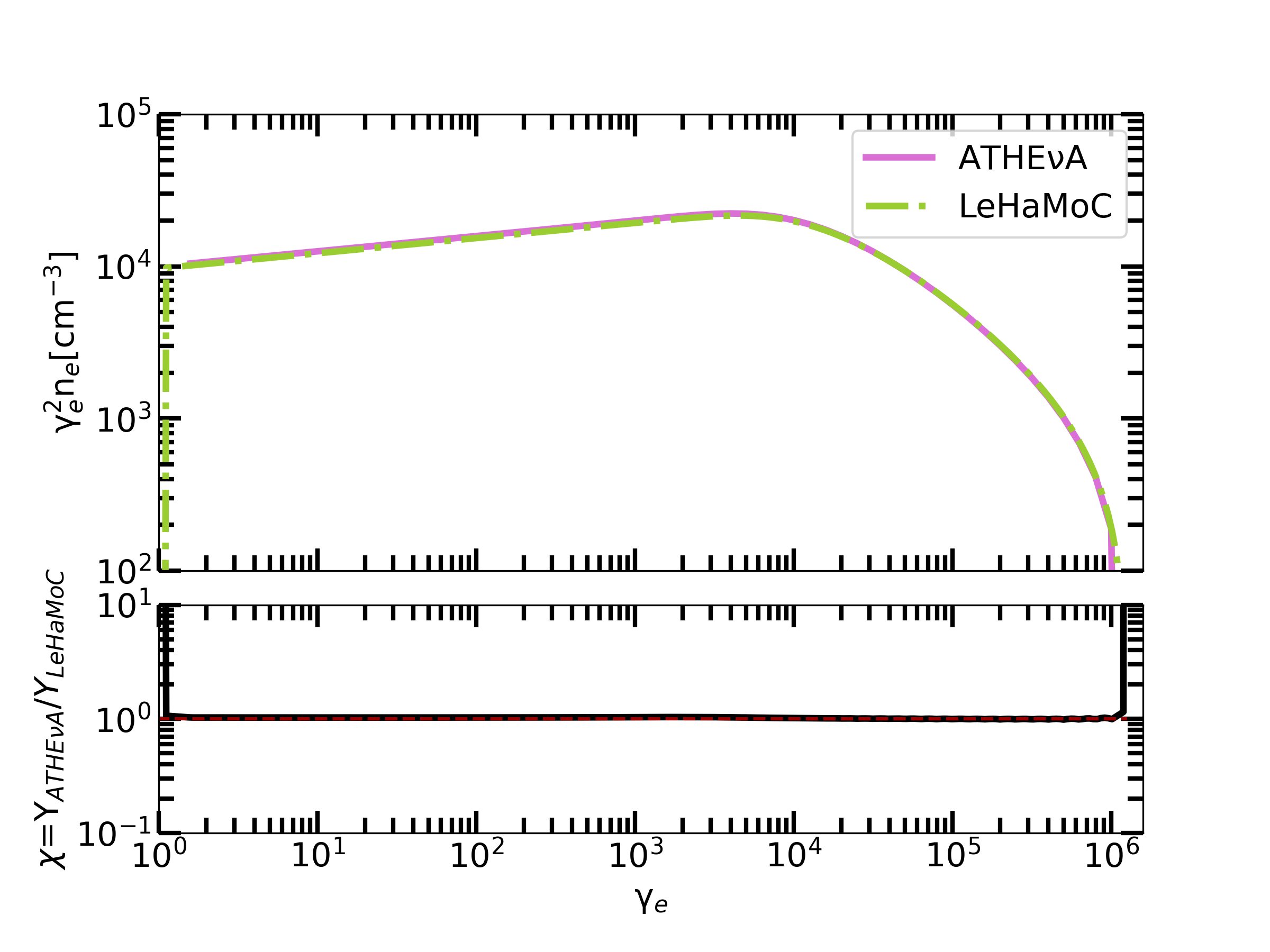}\\
\includegraphics[width=8.5cm]{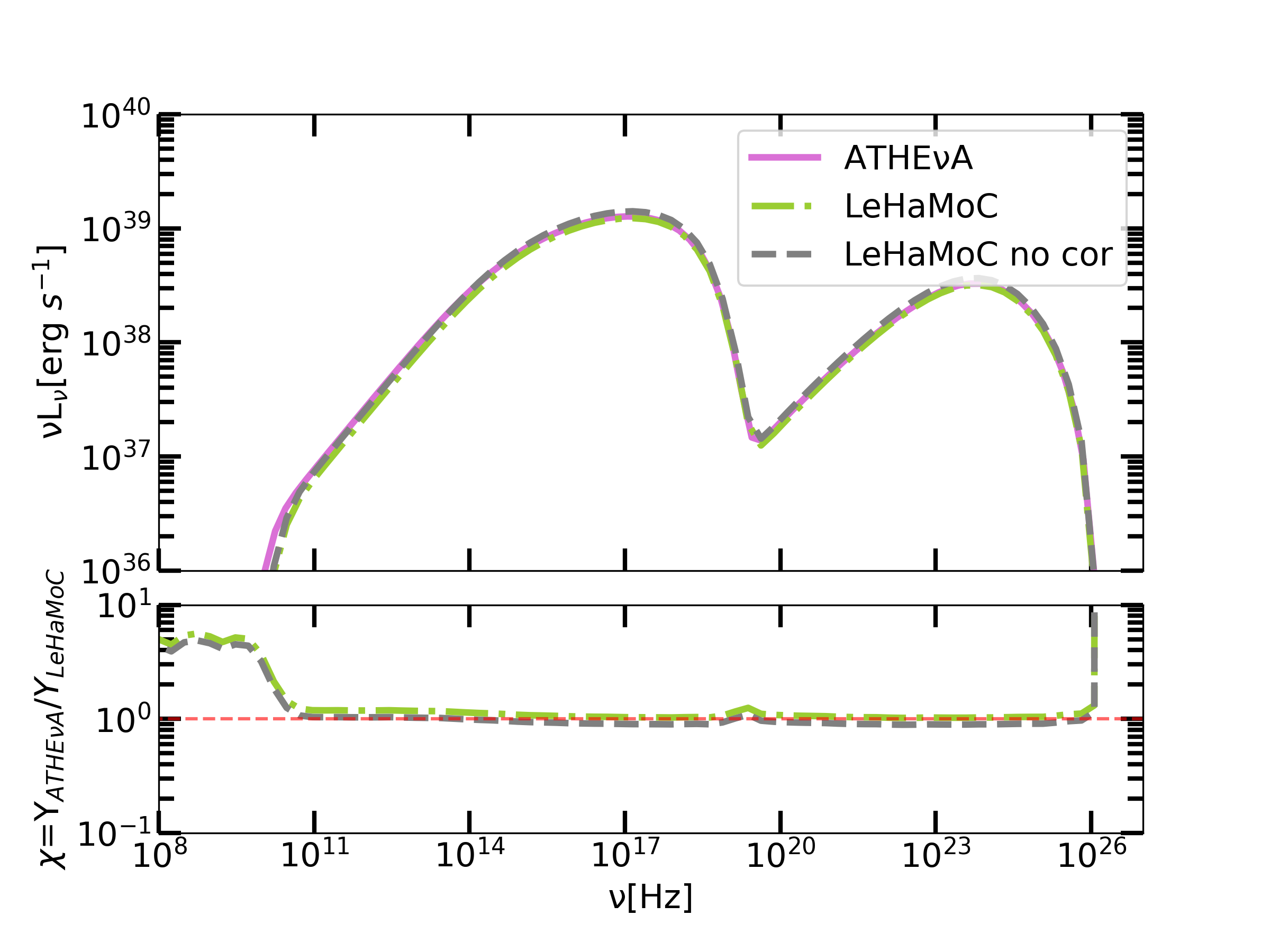}\\
\includegraphics[width=8.5cm]{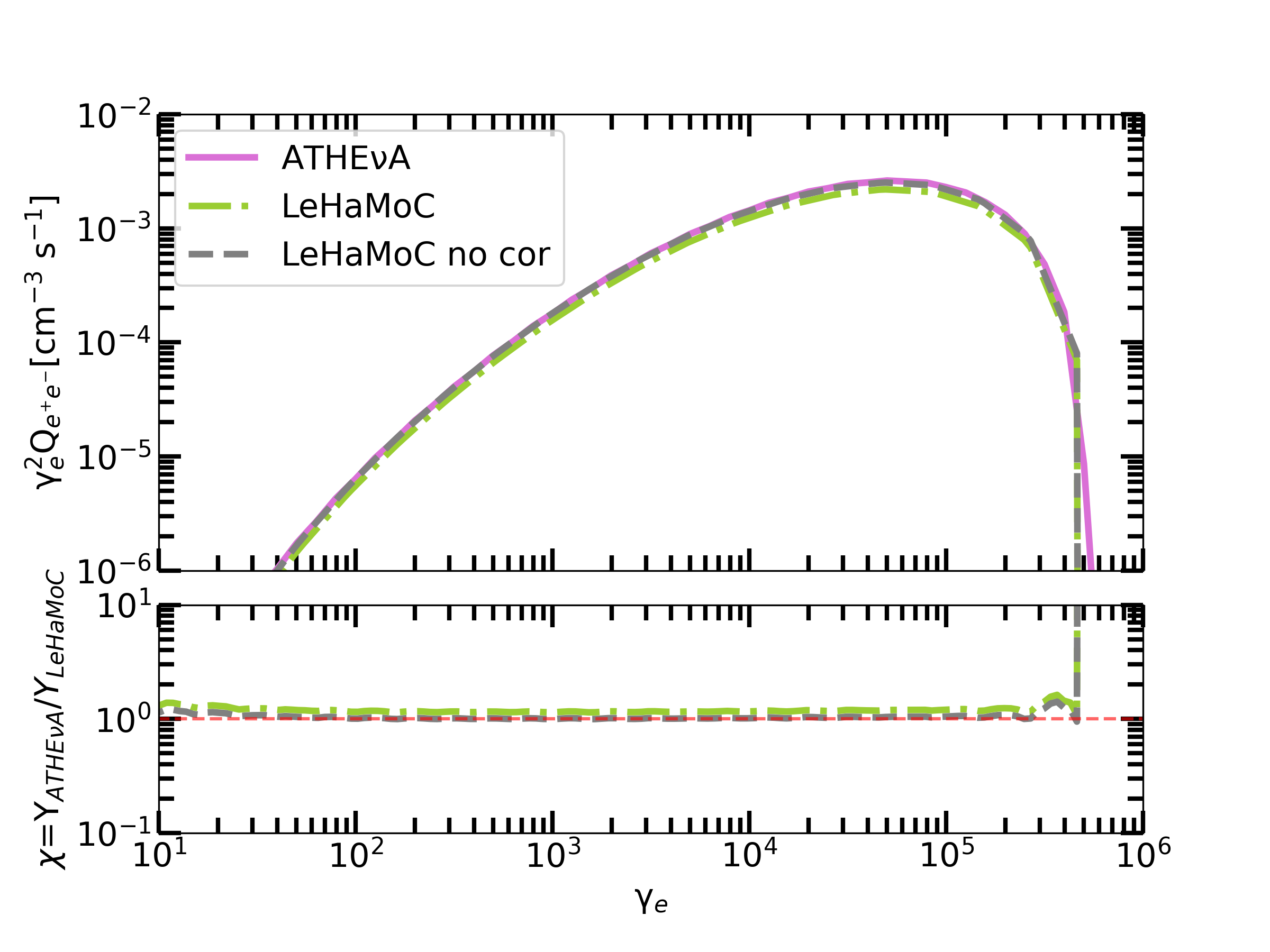}\\
\caption{Steady-state SSC model computed using ATHE$\nu$A and \code \ codes. We show a comparison of (from top to bottom): the steady-state electron spectra, photon spectra, and production rate of pairs due to $\gamma \gamma$ absorption. The bottom panel of each plot shows the ratio $\chi$ of the spectra computed with ATHE$\nu$A and \code. All displayed spectra are measured in the comoving frame of the blob. The parameters used for this test can be found in Table \ref{table1} under the column Test 1. } 
\label{fig:Test1}       
\end{figure}

We present our results in Fig.~\ref{fig:Test1}. The upper panel displays the steady-state electron distributions obtained with the two codes. Both solutions exhibit a smooth break due to synchrotron losses. Both numerical codes have the same location for the cooling break in the electron distribution. Beyond the cooling break, the cooled distribution displays the same power-law behavior as expected by synchrotron theory with a power-law index $s_{e}-1$. In the middle panel, we demonstrate the comoving photon spectra.  We observe that the two codes yield comparable results (see ratio plot) over a broad range of frequencies, both before and after applying the correction for energy balance (see Sec.~\ref{sec:performance}). The two codes predict different locations of the synchrotron self-absorption frequency (0.5 in logarithm), which causes a large difference between the two results at frequencies $\lesssim 100$~GHz. This difference can be attributed to the simplified calculation of the synchrotron self-absorption coefficient in ATHE$\nu$A (see Eq.~40 in \cite{1995A&A...295..613M}); there it is assumed that the electron emits all the power near the critical synchrotron frequency, while in \code \  we use the full relation of the synchrotron power spectrum instead of this approximation (see Eq.~\ref{eq:ssa_coef}). When we implement the delta function approximation we find that the difference is reduced to less than 0.2 in logarithm. Furthermore, we compare the energy spectra of pairs injected via $\gamma \gamma$ pair production in the bottom panel of Fig. \ref{fig:Test1}. The results are in excellent agreement up to the highest Lorentz factor values.

\begin{figure*}
\centering
\includegraphics[width=0.45\textwidth]{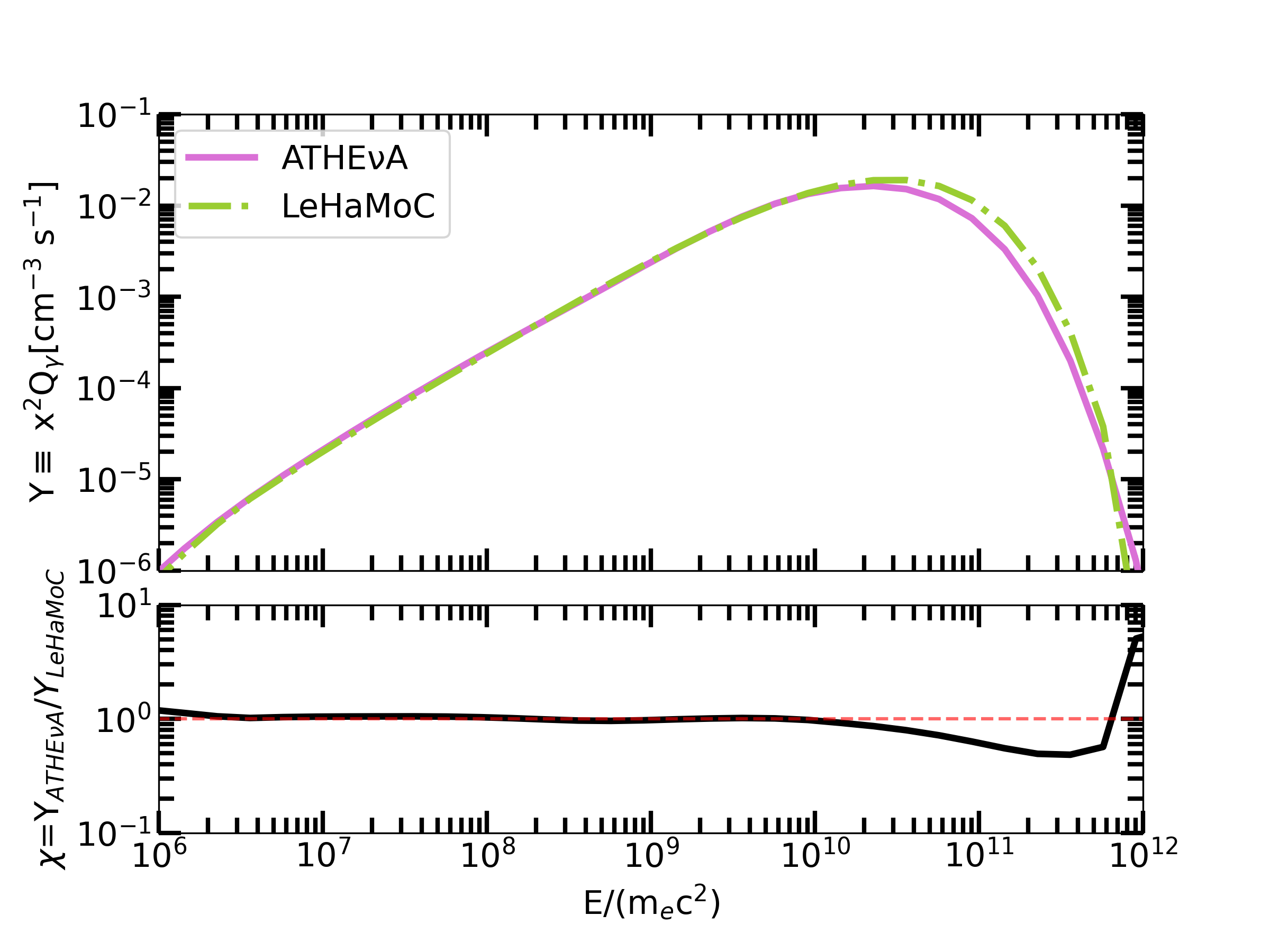}
\includegraphics[width=0.45\textwidth]{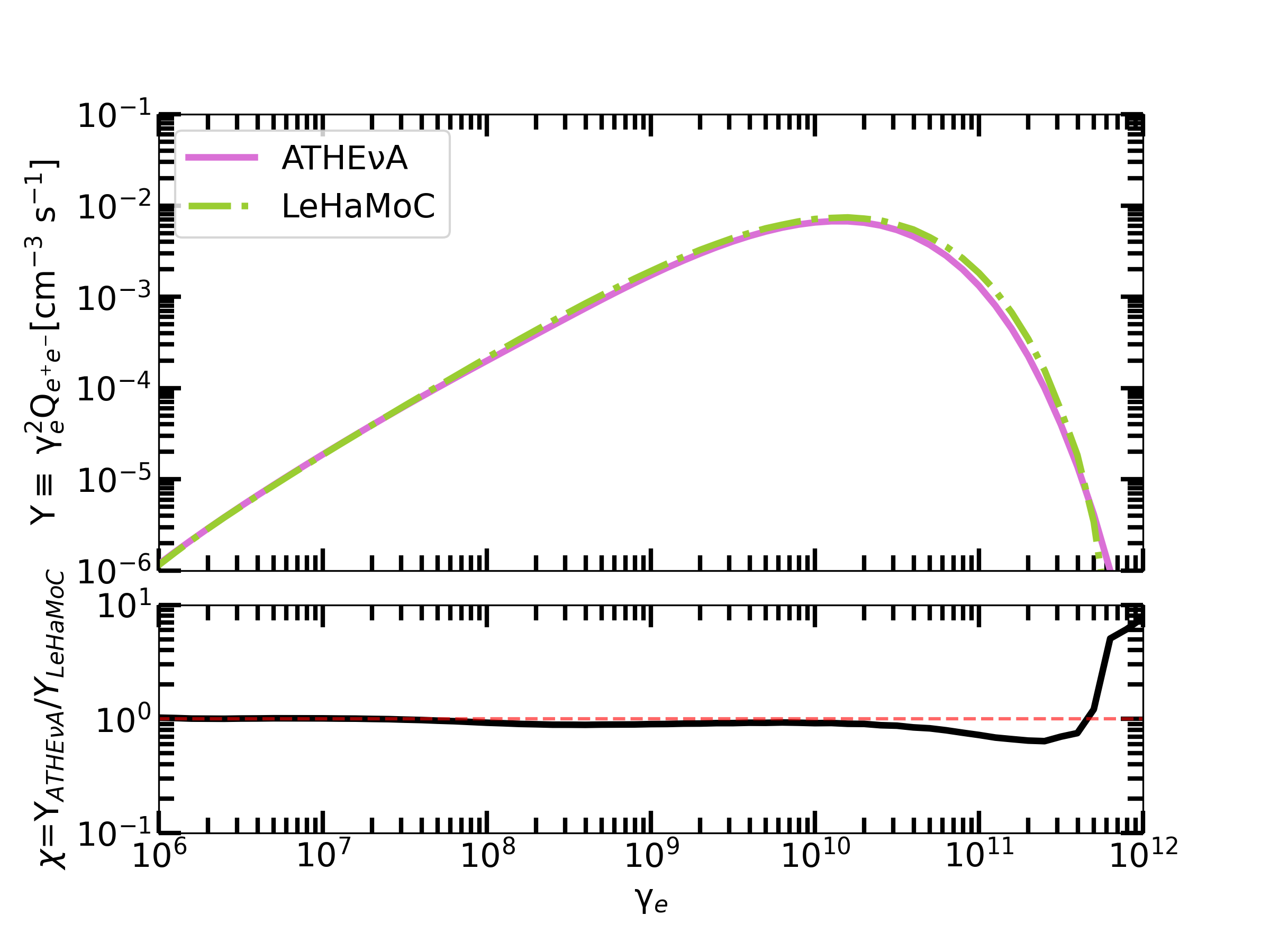}
\includegraphics[width=0.45\textwidth]{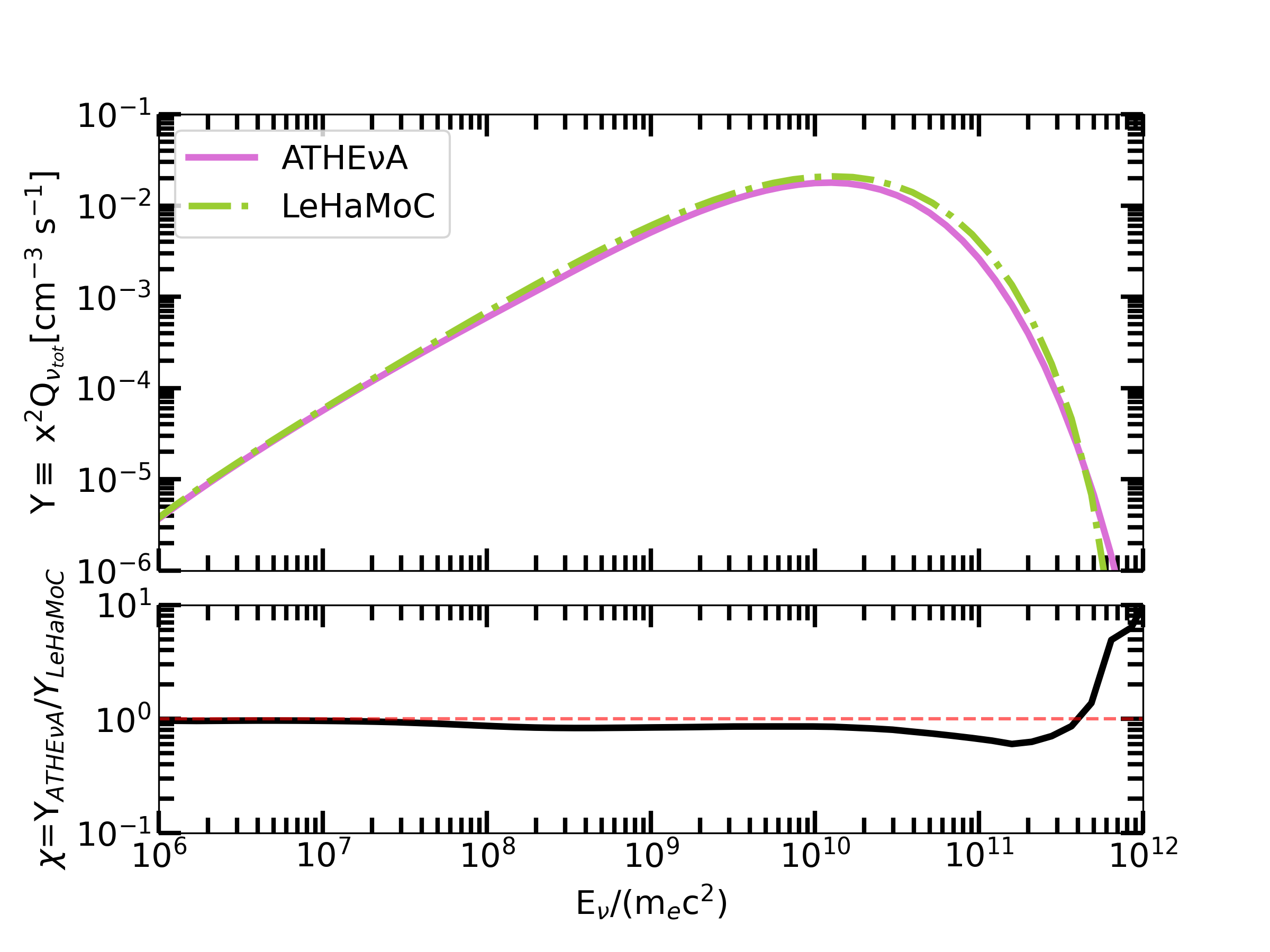}
\includegraphics[width=0.45\textwidth]{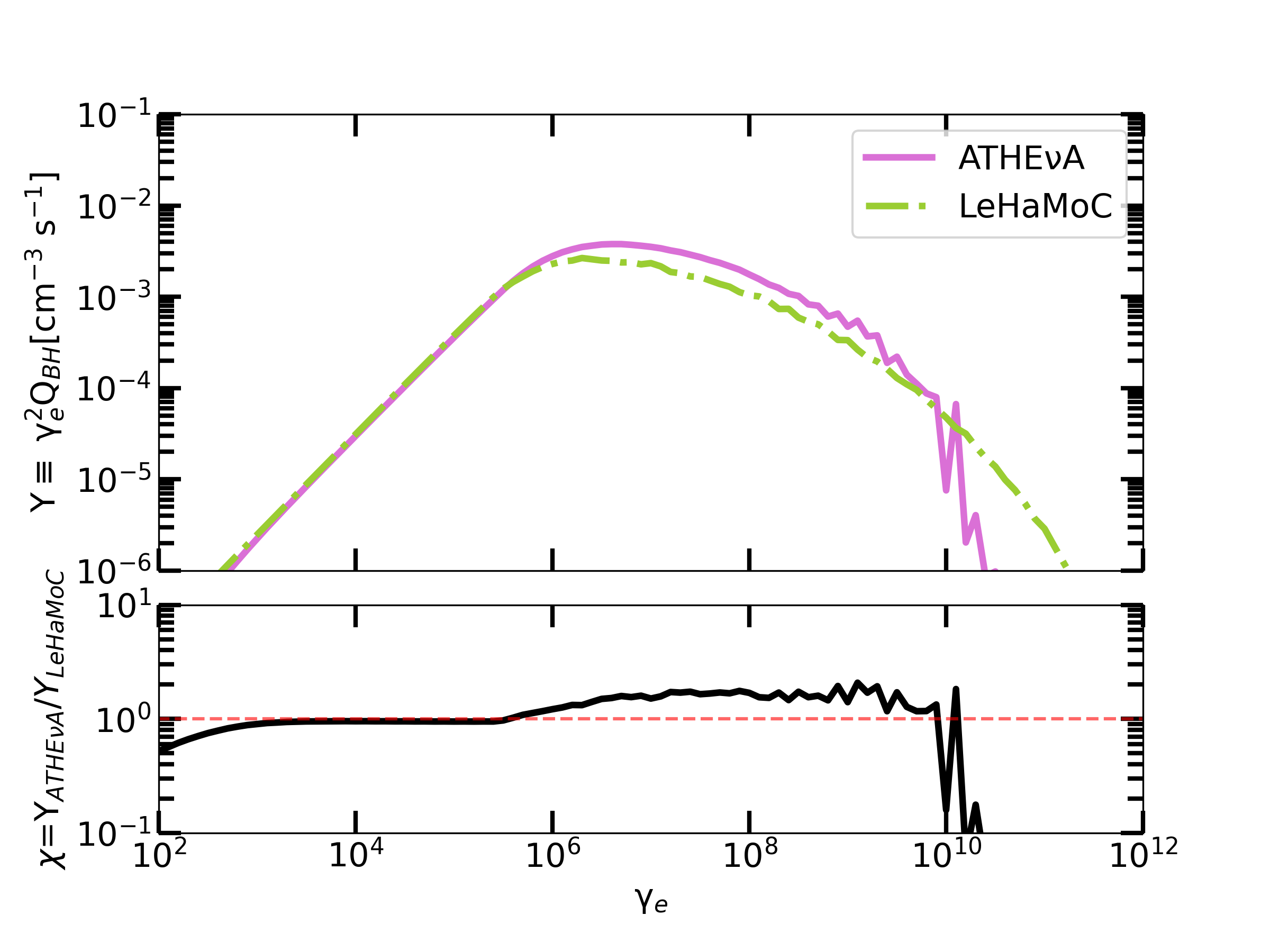}
\caption{Comparison of injection spectra of secondary particles computed using ATHE$\nu$A and \code \ for the case of interactions of relativistic protons with a power-law photon field. From top left and in clockwise order: $\gamma$-ray spectra from neutral pion decay, pair spectra from charged pion decays, all-flavor neutrino spectra from charged pion decays, and pair spectra from the Bethe-Heitler process.
The parameters used for this test can be found in Table \ref{table1} under the column Test 2.}
\label{fig:Test2}       
\end{figure*} 

\subsubsection{Test 2: Production rates of secondaries in proton-photon interactions with extended energy distributions}\label{Test2}
We consider next a generic case for proton-photon interactions where relativistic protons interact with a power-law target photon field. This choice is particularly relevant for astrophysical applications since power-law photon fields are expected to arise from non-thermal emission processes such as synchrotron radiation, and inverse Compton scattering. We are interested in comparing the injection spectra of secondaries produced via Bethe-Heitler and photo-pion production processes on a fixed target photon field. Therefore, we only account for the output of the emissivity functions as described in Appendix \ref{appA} and we neglect all emission produced by the secondaries. We also note that protons do not cool via photo-hadronic interactions due to the low target photon density.

The numerical results obtained with the two codes are shown in Fig.~\ref{fig:Test2}. Starting from the top left figure, we find good agreement in the $\pi^0$ decay $\gamma$-ray emissivity computed with the two codes, as indicated by the ratio $\chi$ in the lower panel; this remains close to unity over a large range of energies, except close to the peak of the spectrum where the difference becomes about 1.4. The pair production spectra from pion decays have a very good agreement (top right panel). The all-flavor neutrino spectra (bottom left panel) have very similar shapes (the ratio is flat across a wide range of energies). Finally, the spectra of Bethe-Heitler pairs are displayed on the bottom right panel. The spectral shapes up to $\gamma_e \sim 10^{10}$ are similar, and there is good agreement in the maximum injected energy (which corresponds to the peak of the spectrum). Nonetheless, the spectrum obtained with ATHE$\nu$A appears to be shifted toward higher Lorentz factors by a factor of $\sim 1.5$. The sharp cutoff of the spectrum at $\gamma_e \sim 10^{10}$ computed with ATHE$\nu$A is artificial because the code considers interaction energies up to $10^4$ above the threshold. 

\subsubsection{Test 3: A time-dependent adiabatic expansion leptonic model}\label{Test3}
We aim to compare the numerical solutions obtained with \code \ with the exact analytical solution of the electron kinetic equation in a time-dependent problem that has no steady-state solution. The primary objective is to investigate the ability of the code to solve time-dependent problems and describe the particle distribution evolution with time. 

For this purpose, we assume an expanding blob where electrons are injected once at the beginning of the simulation with a power-law distribution of slope $s_e$ and are left to cool only via adiabatic losses. The blob has an initial radius $R_0$ and expands with a constant velocity $V_{exp}$. The magnetic field $B_0$ is considered constant throughout the simulation. Under these assumptions, we can obtain an analytical solution for the evolution of the electron distribution that reads \citep{1962SvA.....6..317K},
\begin{eqnarray} 
n_{e}(\gamma_{e},t)&=&\frac{3K}{4\pi R_0^3}(1+\beta_0t)^{-(s_e+2)}\gamma_{e}^{-s_e}\cdot \\ 
& & \Theta\bigg(\gamma_{e,\min}-\frac{\gamma_{e,\min,0}}{1+\beta_0t}\bigg)\Theta\bigg(\frac{\gamma_{e,\max,0}}{1+\beta_0t}-\gamma_{e,\max}\bigg),
\label{eq:N_el_ad} 
\end{eqnarray}
where $\beta_0\equiv V_{exp}/R_0$ and $\Theta(y)$ is the Heavyside function. Moreover, $K$ is a normalization constant, $\gamma_{e,\min}$ and $\gamma_{e,\max}$ are the minimum and maximum Lorentz factor of the electron distribution at time $t$, while $\gamma_{e,\min,0}$ and $\gamma_{e,\max,0}$ are the minimum and maximum Lorentz factors at the moment of injection. 

\begin{figure}
\centering
\includegraphics[width=0.47\textwidth]{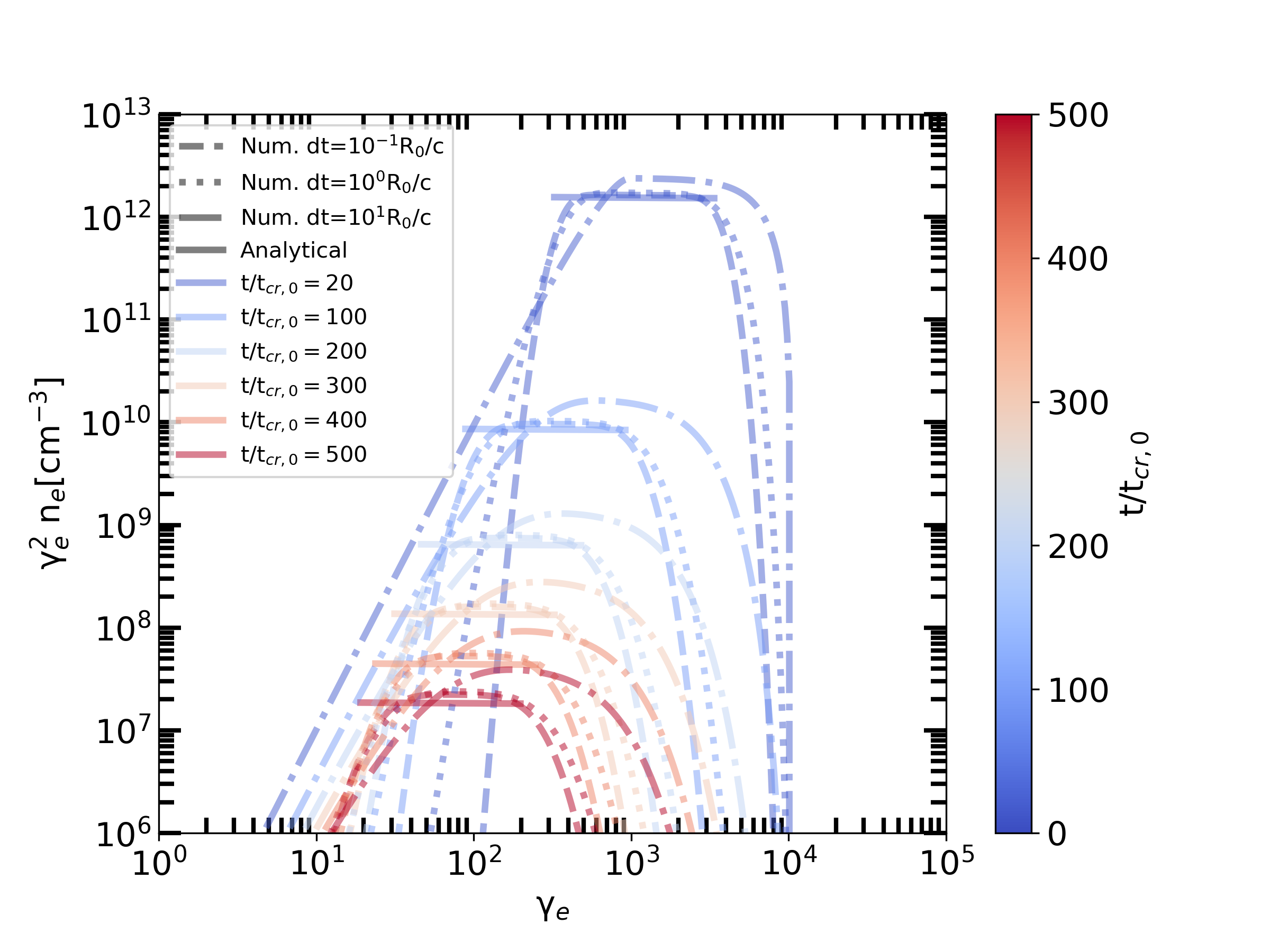}

\caption{Comparison of numerical (dashed, dotted, and dashed-dotted lines) and analytical (solid lines) solutions for adiabatic cooling electrons in an expanding blob with a constant magnetic field. Each numerical result was calculated using a different time step as indicated in the legend. The parameters used for this test can be found in Table \ref{table1} under column Test 3.}
\label{fig:Test3_N_el}       
\end{figure}

Fig.~\ref{fig:Test3_N_el} shows six snapshots of the particle distribution computed analytically using Eq.~\ref{eq:N_el_ad} (solid lines) and numerically with \code.  To highlight the impact of the time step $dt$ on the numerical solutions, we performed runs with different choices (see inset legend). The numerical solution captures the decrease of $\gamma_{e,\min}$ and $\gamma_{e,\max}$ over time due to the adiabatic losses and matches well with the analytical solution in normalization as long as $dt \le R_0/c \equiv t_{cr,0}$. 
In general, choices of $dt > t_{cr,0}$ lead to diffusion of the numerical solution and do not capture the shape of the distribution with precision. 

To further test the accuracy and reliability of our code, we investigate the conservation of electron number throughout the simulation for the numerical results of different time steps. Since electrons are not allowed to escape the system, the total of the electrons,  $N_{e,tot}(t) = V(t) \int n_{e}(\gamma_e,t) \, d\gamma_e$,  should remain constant over time. 

\begin{figure}
\centering
\includegraphics[width=0.47\textwidth]{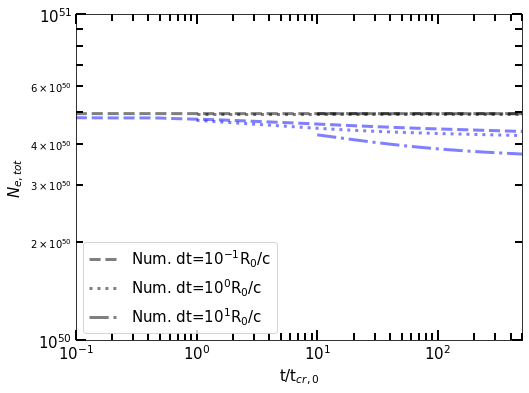}
\caption{Evolution of the total number of particles $N_e$ with time for the same case shown in Fig~\ref{fig:Test3_N_el}. Different types of lines show $N_e$ as derived from the numerical solutions for different choices of the time step (see inset legend). The black lines represent the number of particles calculated by considering all grid points, while the blue lines represent the integration of the particle distribution within the analytically expected range of Lorentz factors, i.e. $\gamma_{e,\min}(t)$ and $\gamma_{e,\max}(t)$. For comparison, the analytical solution yields $N_{e,tot}=4.9\times10^{50}$ electrons.}
\label{fig:Test_3_N_e}   
\end{figure}

We find that $N_{e,tot}$ remains constant for all values of $dt$ as long as the integration is performed over a wide range of Lorentz factors (see black lines in Fig.\ref{fig:Test_3_N_e}). If, however, the integration is limited between the analytically expected $\gamma_{e, \min}(t)$ and $\gamma_{e,\max}(t)$, we find particle conservation for $dt < R_0/c$ as demonstrated by the blue lines in Fig.~\ref{fig:Test_3_N_e}). We further comment on the impact of $dt$ on the numerical solution of dynamically involving systems through another illustrative example in Appendix~\ref{appB}.

By utilizing the analytical solution given by Eq.~\ref{eq:N_el_ad} for an expanding blob subject to adiabatic losses, we can also determine the evolution of the synchrotron-self absorption frequency $\nu_{ssa}$ over time,
assuming that synchrotron radiation is the only emission mechanism for electrons \citep[see also][for leptonic models with expansion]{2022A&A...657A..20B}. We find that

\begin{equation}
\nu_{ssa}(t)\propto (1+\beta_0t)^{-\frac{2(1+s_e)}{s_e+4}}\stackrel{\text{$\beta_0t\gg1$}}{\propto}t^{-\frac{2(1+s_e)}{s_e+4}}.
\label{eq:nu_ssa_1}
\end{equation}

\begin{figure}
\centering
\includegraphics[width=0.47\textwidth]{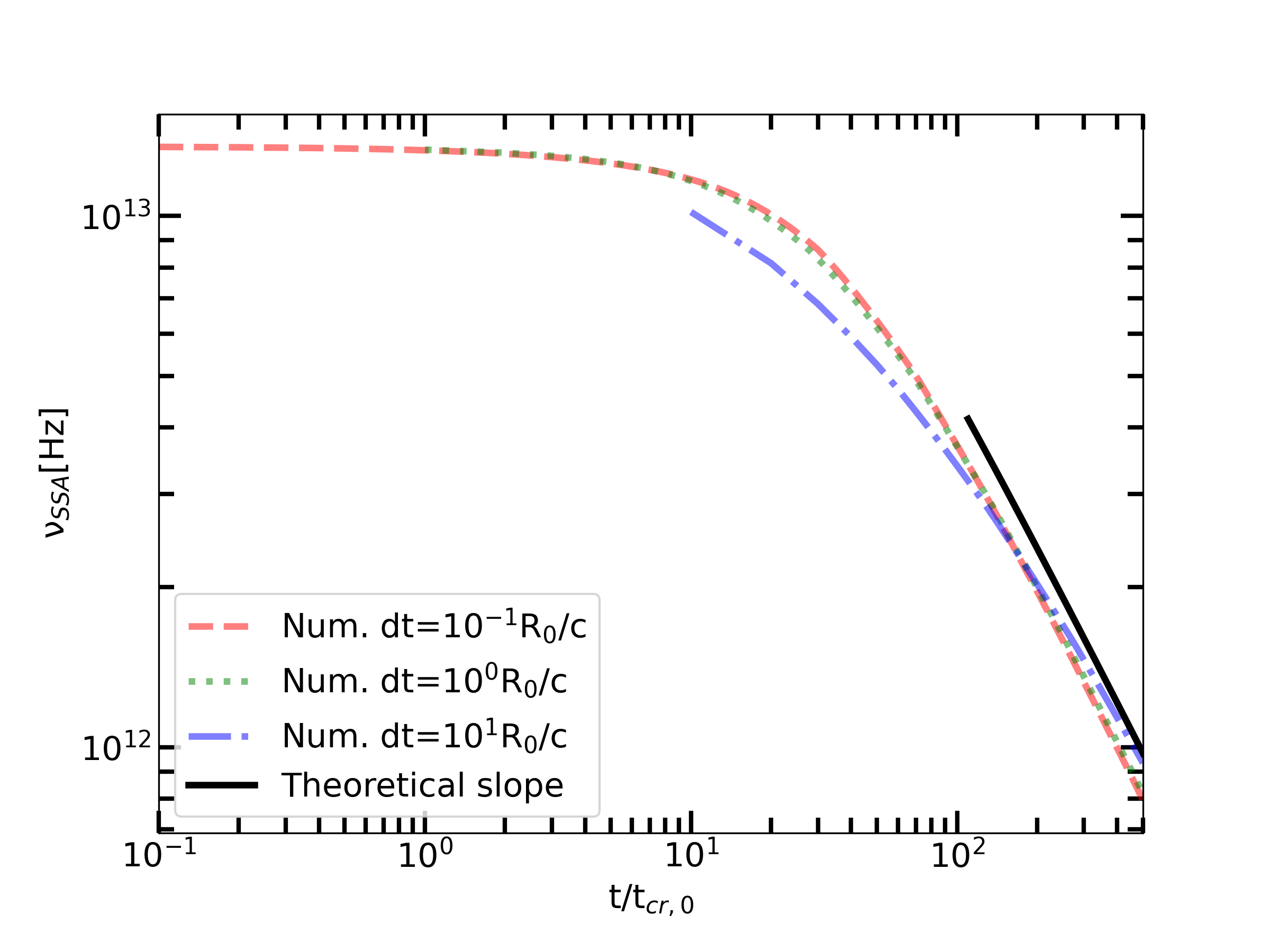}
\caption{Evolution of the synchrotron-self absorption frequency $\nu_{ssa}$ with time for the same case shown in Fig~\ref{fig:Test3_N_el}. Different types of lines show $\nu_{ssa}$ as derived from the numerical solutions for different choices of the time step (see inset legend). For comparison, we show the analytical prediction for the late-time ($t \gg R_0/V_{exp}$) evolution of $\nu_{ssa}$ with time.}
\label{fig:Test_3_nu_ssa}   
\end{figure}

We compare the numerical results of $\nu_{ssa}$\footnote{We determine the ssa frequency by numerically solving the equation, $\tau_{\nu_{ssa}}=\alpha_{\nu_{ssa}}R\simeq 1$, where $\tau_{\nu_{ssa}}$ is the ssa optical depth, $\alpha_{\nu_{ssa}}$ is the absorption coefficient at this frequency and $R$ is source radius (i.e. the distance traveled by the photon inside the source).} with the analytical scaling relation in Fig. \ref{fig:Test_3_nu_ssa}. The solid line has the same slope as the one predicted analytically for $\beta_0t\gg1$, while the numerical results for different time steps are depicted with dashed, solid, and dashed-dotted lines. We observe that for time steps $dt\le t_{cr,0}$, the evolution of numerically derived $\nu_{ssa}$ follows the expected time dependency at late times. However, for $dt > t_{cr,0}$, the numerical evolution of $\nu_{ssa}$ with time is slower than predicted (see difference in slopes).

\subsubsection{Test 4: A steady-state leptohadronic model}\label{Test4}
Finally, we present a generic steady-state leptohadronic scenario. The objective of this test is to compare the broadband photon and neutrino spectra obtained from ATHE$\nu$A and \code, considering the inclusion of all physical processes (except for pp collisions, which are not included in ATHE$\nu$A). To initiate the simulation we assume that relativistic electrons and protons are injected with a distribution $dN_i/d\gamma=K\gamma^{-s_i}e^{-\gamma/\gamma_{i, \rm coff}}$, for $\gamma \ge \gamma_{i, \min}$, into a spherical blob with radius $R_0$ and magnetic field strength $B_0$. Moreover, relativistic pairs are injected in the source through Bethe-Heitler pair production, pion production through proton photon interactions, and pair creation via $\gamma\gamma$ interactions. Particles and photons are allowed to escape on a timescale equal to the light-crossing time of the source, i.e. $t_{cr} = R_0/c$. The parameter values, which are presented in Table~\ref{table1}, are one of the parameter sets used in the Hadronic Code Comparison Project (Cerruti et al., in prep.) where four radiative transfer codes including ATHE$\nu$A are compared.

\begin{figure}
\centering
\includegraphics[width=0.47\textwidth]{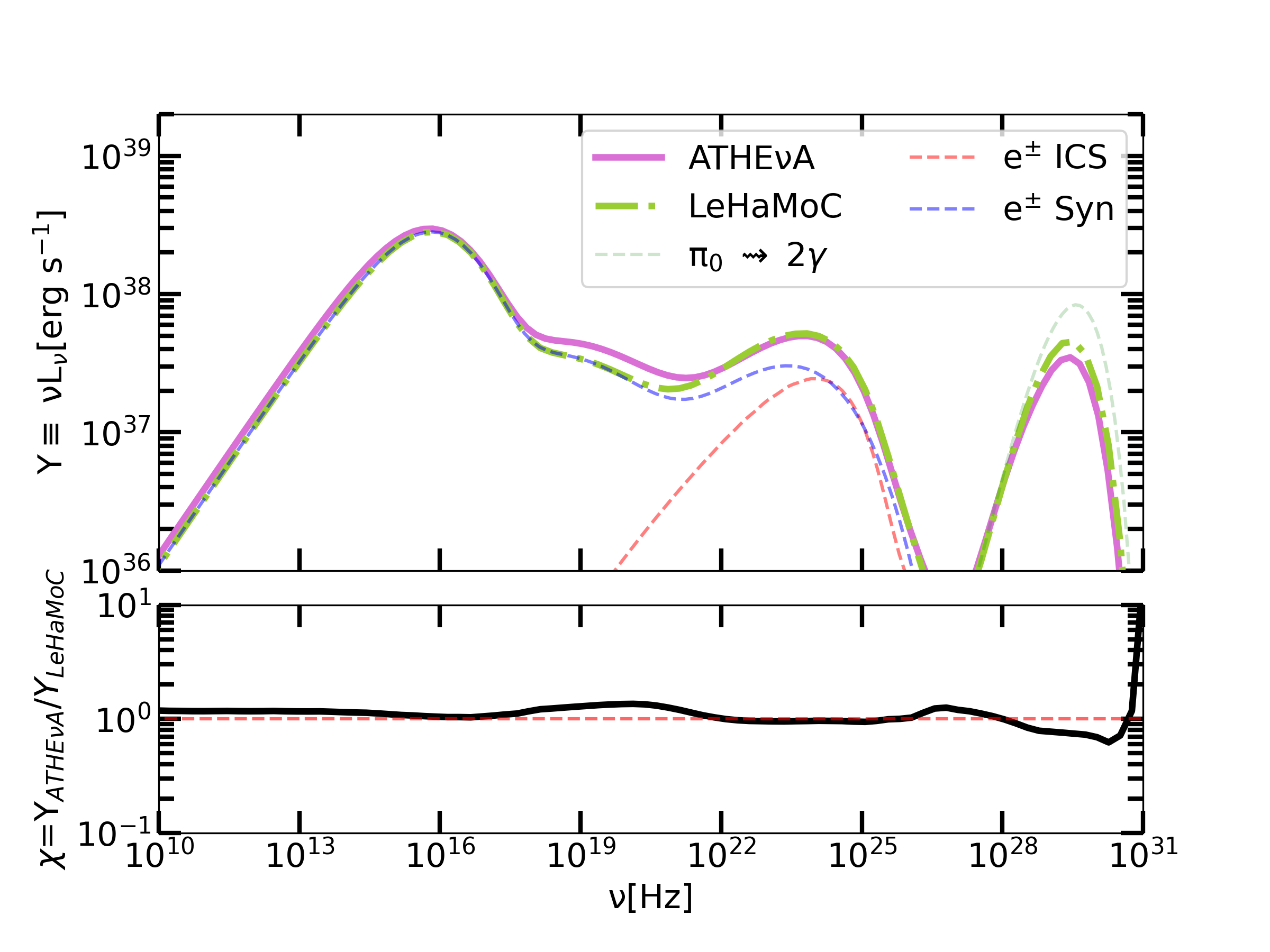}

\caption{Comparison of the steady-state photon spectra of ATHE$\nu$A code and \code \, for a leptohadronic scenario for blazar emission. All displayed spectra are measured in the comoving
frame of the spherical blob. The colored dashed lines are used to indicate the contribution to the photon spectrum coming from the pairs (blue and red) and the decay of neutral pions (green) to $\gamma$-rays before attenuation. The parameters used for this test can be found in Table~\ref{table1} under column Test 4.}
\label{fig:Test4}       
\end{figure}

Both codes yield remarkably similar broadband photon spectra (thick lines), as shown in the top panel in Fig.~\ref{fig:Test4} -- see also the ratio plot in the bottom panel. We also present a decomposition of the total photon emission computed with \code\, into different spectral components. For the selected parameter values the electron synchrotron spectrum contributes significantly to the X-ray range (up to $\sim 10^{17}$ Hz). These synchrotron photons are energetic enough and can serve as targets for both Bethe-Heitler and pion production interactions with the protons in the system. Both interactions lead to the injection of high-energy pairs, with extended energy distributions (see e.g. Fig.~\ref{fig:Test2}) that will radiate via synchrotron and inverse Compton processes. In particular, the broad component peaking at approximately $\sim 10^{19}$~Hz corresponds to the emission of synchrotron-cooled Bethe-Heitler pairs~\citep[see also][]{Petro2015}. 

Additionally, the component peaking at approximately $\sim 10^{24}$ Hz arises from the synchrotron emission of pairs injected via charged pion decays. Inspection of the ratio plot suggests that both codes produce similar results across a wide range of photon energies where the secondary pair emission dominates. Finally, the spectrum exhibits a peak resulting from the direct decay of $\pi^0$ mesons, which generates very high-energy $\gamma$-rays ($\sim 10^{29}$ Hz). The observed decrease in luminosity (compare thick and thin green lines), resulting from the absorption of high-energy $\gamma$-ray photons by GHz photons, is consistently reproduced by both codes. Similarly, there is a good agreement in the all-flavor neutrino spectra computed with both codes. It is noteworthy though that the implementation used in \code \, speeds up the computation of the multi-messenger emission by a factor of $\sim 3$ compared to ATHE$\nu$A, bringing it down to $\sim 12$ min. An even faster computation is also possible but at the cost of reduced accuracy.

\begin{figure}
\centering
\includegraphics[width=0.47\textwidth]{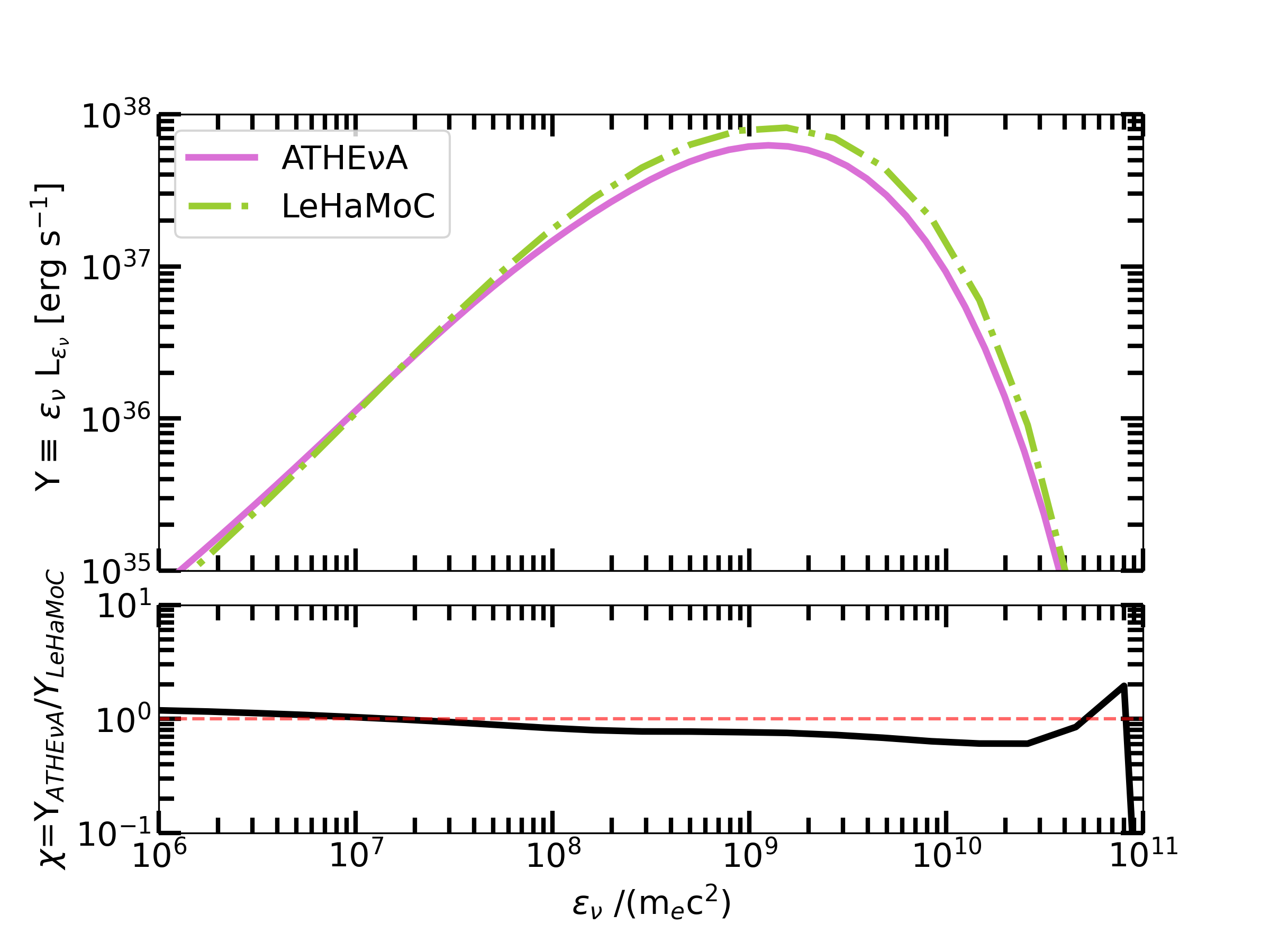}
\caption{Comparison of the steady-state all-flavor neutrino spectra computed with ATHE$\nu$A and \code \, codes for Test 4. }
\label{fig:Test4_neutrino_spec}       
\end{figure}

\renewcommand{\arraystretch}{1.2}
\begin{table}[]
    \centering
\caption{Parameter values for scenarios used to test the accuracy and performance of \code.}
    \begin{threeparttable}
    \begin{tabular}{@{}lcccc@{}}
\hline
Parameters & Test 1 & Test 2 & Test 3 & Test 4 \\
\hline
$R_0$ [cm]  & $10^{15}$ & $10^{15}$ & $10^{13}$ & $10^{16}$ \\
$B_0$ [G]  & 1 & 10 & 10 & 0.1 \\
$V_{exp}/c$ & 0 & 0 & 0.1 & 0   \\
$\gamma_{e,\min}$    & 1 & - & $10^{3}$ & $10^{0.1}$  \\
$\gamma_{e,\rm coff}$\tnote{*}  & - & - & - &  $10^{5.5}$   \\
$\gamma_{e,\max}$   & $10^6$ & $10^{4}$ & $10^{4}$ & $10^{11}$  \\
$\gamma_{p,\min}$    & - & 1 & - & $10^{0.1}$ \\
$\gamma_{p,\rm coff}$\tnote{*}  & - & $10^8$ & - & $10^{6.2}$ \\
$\gamma_{p,\max}$  & - & $10^9$ & - &  $10^{7}$ \\
$s_e$ & 1.9 & - & 2.01 & 2.01   \\
$s_p$ & - & 1.9 & - & 2.01  \\
$L^{inj}_e~[\rm erg \ s^{-1}]$  & $3.1\cdot10^{40}$ & - & $10^{48}$ & $3.7 \cdot 10^{40}$  \\
$L^{inj}_p~[\rm erg \ s^{-1}]$  & - & $1.1\cdot10^{45}$ & - & $2.8 \cdot 10^{46}$  \\
$U_{ext}~[\rm erg \ s^{-1}]$  & - & $3.6 \cdot 10^{-2}$ & - & -   \\
$\epsilon^{\min}_{ext}~[\rm erg]$ & - &  $8.2\cdot 10^{-13}$ & - & -   \\
$\epsilon^{\max}_{ext}~[\rm erg]$ & - &  $8.2 \cdot 10^{-8}$ & - & -   \\
Photon Index & - & 2 & - & -   \\
\hline
    \end{tabular}
    \begin{tablenotes}
        \item[*] Particle distribution is modeled as $N_i(\gamma)=K\gamma^{-s_i}e^{-\gamma/\gamma_{i, \rm coff}}$, for $\gamma \ge \gamma_{i, \min}$. 
    \end{tablenotes}
    \label{table1}
    \end{threeparttable}
\end{table}

\section{Astrophysical applications}\label{sec:applications}
To demonstrate the fitting capabilities of the newly developed code we first model a blazar SED using {\tt emcee} \citep{2013PASP..125..306F}, a python implementation of the Affine invariant Markov chain Monte Carlo (MCMC) ensemble sampler. This allows us to better estimate the uncertainties in model parameters and to explore possible degeneracies in this multi-parameter problem. To illustrate the capabilities in computing hadronic-initiated electromagnetic cascade spectra,  we present a pp-interaction model for the neutrino signal from the nearby Seyfert galaxy NGC 1068. 

\subsection{Blazar SED fitting}\label{sec:blazar-sed}
For this application, we choose \hsp \,  \citep{Giommi2020, 2020ApJ...902...29P}, which is a BL Lac object at redshift $z=0.557$ \citep{10.1093/mnrasl/slaa056, 2020ApJ...902...29P} that belongs to the rare class of extreme blazars\footnote{The low-energy component of these blazars typically peaks (in a $\nu F_\nu$ versus $\nu$ plot) at frequencies exceeding $10^{16}$ Hz.} \citep{Biteau2020}. This blazar has been detected in GeV $\gamma$-rays by the \fermi \, Large Area Telescope (LAT) and is part of the  4FGL catalog \citep{2020ApJS..247...33A}. It has also been possibly associated with a high-energy neutrino IceCube-200107A detected by the IceCube Neutrino Observatory \citep{2020GCN.26655....1I} one day before the blazar was found to undergo a hard X-ray flare. 

Various emission models of the 3-day X-ray flare, including one-zone leptohadronic scenarios, were presented in \citet{Petro2020}. In all cases, the parameter values were selected based on physical arguments instead of being derived from a fitting procedure. The long execution time of the radiative code ATHE$\nu$A, even for a pure leptonic emission model, deemed necessary this ``eyeball'' description of the SED~\citep[see also][]{Petro2015, Petro2017}. Here, we demonstrate that SED fitting using the newly developed code and a Bayesian inference approach is feasible with limited computational resources. \hsp \, is ideal for this application. First, our results can be compared to SED models that are available in the literature. Second, we can investigate how the different spectral coverage of the low- and high-energy SED humps (see below for more details) affects our results.

\begin{figure}
\centering
\includegraphics[width=0.45\textwidth]{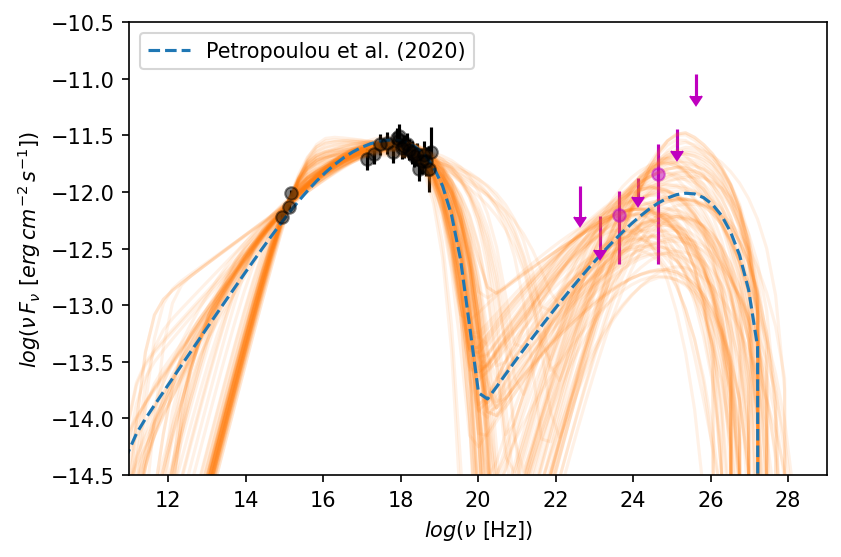} 
\caption{SED of \hsp \ using data from \citet{Giommi2020}. Black symbols indicate observations taken on January 11, 2020 (i.e. soon after the arrival of the neutrino alert). Magenta symbols show the time-integrated \fermi-LAT data over a period of 250 days prior to the neutrino alert. SSC spectra computed for a random sample of 100 points from the posterior distributions are overplotted (solid orange lines). For comparison, the SSC model from \citet{Petro2020} is also shown (dashed blue line). Photon attenuation by the extragalactic background light (EBL) is not taken into account.}
\label{fig:sed-ssc}
\end{figure}

We use the multi-wavelength data of  January 11, 2020 (4 days after the neutrino detection) when the source was observed for the first time in hard X-rays by \nustar~\citep{nustar2013}. The quasi-simultaneous observations in the UV, soft X-rays, and hard X-rays provide a detailed picture of the low-energy part of the spectrum (see black symbols in Fig.~\ref{fig:sed-ssc}). On the contrary, the high-energy part of the spectrum is less constrained observationally. Given that \hsp \,  is a faint $\gamma$-ray source, the \fermi \, data need to be averaged over a long period (250 days) to allow for a significant detection of the source at least in two energy bins (see magenta symbols in Fig.~\ref{fig:sed-ssc}). For details about the data selection and analysis, we refer the reader to \citet{Giommi2020}.  

The statistical uncertainties of the flux measurements typically underestimate uncertainties that stem from the non-simultaneity of data used in the SED, since blazar emission is variable on timescales of days to months (e.g. the X-ray data are obtained from $\sim$ks observations, while the $\gamma$-ray data are averaged over 250 days). We, therefore, add a term $\ln{f}$ to the likelihood function to account for any source of uncertainty that is not included in the statistical uncertainties of the measurements \citep[for a similar application, see][]{2023MNRAS.520..281K}, 
\begin{equation}
\ln{\mathcal{L}} =  -\frac{1}{2} \sum_{i}^{} \frac{(f_{m,i}-f_i)^2}{\sigma_{\rm tot, i}^2}+e^{\sigma_{\rm tot, i}^2}. 
\end{equation}
Here, $f_{m,i}$ stands for the model flux (in logarithm) evaluated at the frequency of the $i$-th data point, $f_i$ is the measured flux, and $\sigma_{\rm tot, i}$ is the total variance, which is defined as 
\begin{equation}
\sigma_{\rm tot, \rm i}^2 = \sigma_{\rm i}^2 + e^{2\ln{f}}, 
\end{equation}
with $\sigma_{\rm i}$ being the errors of the flux measurements. We also note that the upper limits are included in fit by adding the following term in the likelihood function \citep[see][]{2012PASP..124.1208S,2020MNRAS.491..740Y}:
\begin{equation}
\sum_j^{} \ln \left ( \sqrt{\frac{\pi}{2}}\sigma_j \left [1+\textrm{erf}\left(\frac{f_{lim,j}-f_{m,j}}{\sqrt{2} \sigma_j}\right) \right] \right),
\end{equation}
where $f_{lim, j}$ is the $j$-th flux upper limit (in logarithm) and $\sigma_j$ is the respective uncertainty.

We apply first an SSC model with synchrotron self-absorption and intrinsic $\gamma \gamma$ pair production included. The electron distribution is modeled as a power law of slope $s_e$ that starts from $\gamma_{e,\min}$ and extends to a sharp high-energy cutoff at $\gamma_{e, \max}$. In the MCMC fitting all parameters are sampled from uniform distributions in logarithmic space, except for the power-law slope of the particle distribution, which is sampled from a uniform distribution in linear space. We produce a chain with 48 walkers that are propagated  50,000 steps each and discard the first 5,000 steps of each chain as burn-in. The corner plot showing the posterior distributions of the parameters is presented in Fig.~\ref{fig:corner-ssc}, and a random selection of SSC spectra computed using 100 posterior values is shown in Fig.~\ref{fig:sed-ssc}. For comparison purposes, we include an SSC spectrum used in \citet{Petro2020} as an ``eyeball'' description of the SED (model D). The displayed SSC spectra highlight the importance of obtaining flux measurements with small uncertainties. In particular, the UV and X-ray measurements act as nodes for the model and help decrease the spread in the predicted synchrotron component of the SED. On the contrary, the high-energy hump of the SED is less constrained (due to upper limits and large error bars in the LAT energy range), leading to a larger spread in SSC spectra (and physical conditions) that are statistically compatible with the observations.  While model D of \citet{Petro2020}  falls within the range of SSC spectra found by the Bayesian inference, it is only one of the possible realizations of the emitting region (see also corner plot in Fig.~\ref{fig:corner-ssc}).

\begin{figure}
\centering
\includegraphics[width=0.45\textwidth]{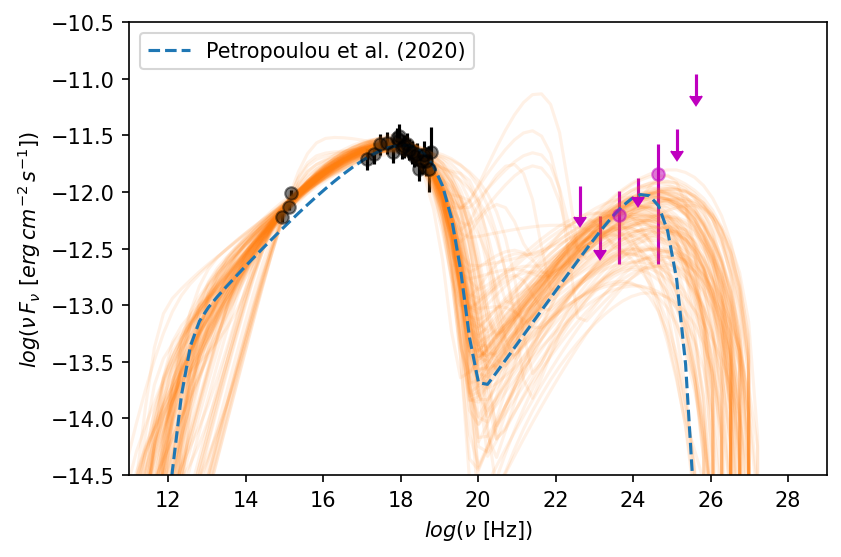} 
\caption{Same as in Fig.~\ref{fig:sed-ssc} but for a proton synchrotron (PS) model of the SED. For comparison, the PS model from \citet{Petro2020} is also shown (dashed blue line).}
\label{fig:sed-psyn}
\end{figure}

We next consider the simplest lepto-hadronic scenario where relativistic protons are injected into the source with a power-law distribution and produce photons only via synchrotron radiation. This is known as the proton-synchrotron (PS) model for $\gamma$-ray blazar emission~\citep{Aharonian2000, Muecke2001}. For this illustrative example, we neglect photo-hadronic interactions but note that these may not be negligible for all the parameters explored by the walkers in the chain. We defer a full lepto-hadronic MCMC modeling for future work. Fig.~\ref{fig:sed-psyn} shows 100 SEDs (solid orange lines) obtained by randomly selecting 100 parameter sets from the posterior distributions shown in Fig.~\ref{fig:corner-psyn}. The proton-synchrotron model discussed in \citet{Petro2020} is also plotted for comparison purposes (dashed blue line). Without imposing any constraints on the parameters (e.g. same power-law slopes or maximum energies for the electron and proton distributions), the low- and high-energy humps of the SED are decoupled. This, in combination with the $\gamma$-ray upper limits,  leads to solutions with a spread in the predicted $\gamma$-ray peak energy. This spread in solutions is also reflected in the broad posterior distributions of the parameters describing the proton distribution. For example, the minimum proton energy and power-law slope cannot be constrained, thus justifying the simplifying assumption of fixing these parameters to default values~\citep[see][]{Petro2020}.

\subsection{Electromagnetic cascades in high-energy neutrino sources}
NGC 1068 is a prototypical Seyfert 2 galaxy at $d_L \simeq 14.4$~Mpc\footnote{It is important to acknowledge that there is a degree of uncertainty regarding the values, which varies between 10.6 $\pm$ 3 Mpc (see references at NASA/IPAC Extragalactic Database.)} \citep{2004MNRAS.350.1195M} exhibiting AGN and starburst activity in its central region \citep{1994ApJ...430..545N}. Due to its proximity, it is the best-studied galaxy of this type. 
The estimated mass of the supermassive black hole (SMBH) in NGC 1068 is approximately 20 million times the mass of the Sun, i.e. $M \simeq 10^{7.3} M_{\odot}$ \citep{2006A&A...455..173P}, and the corresponding Schwarzschild radius is $R_s \simeq 6 \times 10^{12}$ cm.  It is a well-known bright X-ray source with estimated intrinsic X-ray luminosity (in the 0.1-200 keV range) of $\sim 4.6\times10^{43}$~erg s$^{-1}$~\citep{2015ApJ...812..116B}. NGC 1068 is also detected in GeV $\gamma$-rays by \fermi-LAT \citep{2010A&A...524A..72L,2020ApJS..247...33A}, but upper limits are set at even higher energy by MAGIC \citep{2019ApJ...883..135A}. Recently, the IceCube Collaboration has reported a high-energy neutrino excess (at 4.2 $\sigma$ confidence level) associated with NGC 1068 \citep{2022Sci...378..538I}, strengthening previous reports of a $2.9\sigma$ excess in the 10-year time-integrated search \citep{2020PhRvL.124e1103A}.

Many models for the neutrino and/or $\gamma$-ray emission of NGC~068 can be found in the literature \citep[e.g.][]{Murase_2020, 2021ApJ...922...45K, 2022arXiv220702097I, 2022ApJ...939...43E}. We do not attempt to offer a new interpretation but rather present the capabilities of our code using one of the published one-zone models for NGC 1068. In particular, we adopt the AGN corona scenario of \citet{2022ApJ...941L..17M} in which neutrinos are produced via pp interactions in a magnetized corona and $\gamma$-rays are attenuated via the disk-corona radiation field. For simplicity, we ignore the emission from accelerated (primary) electrons in the corona and focus on the electromagnetic cascade developed by secondary pairs. While details of the model can be found in \citet{2022ApJ...941L..17M}, we outline below the basic model ingredients for completeness.

We assume that protons are accelerated in the corona into a power-law distribution, i.e. $dN/dE_p \propto E_p^{-s_p}$, $E_p \ge E_{p,\min}$, where $E_p = m_p \gamma_p c^2$. Relativistic protons can interact with the cold (i.e. non-relativistic) protons present in the corona. The number density of the latter is approximated as $n_{p,c} \sim \sqrt{3} \tau_T/(\zeta_e \sigma_T R)$, where $\tau_T$ represents the coronal optical depth to Thomson scattering, $\zeta_e$ is the pair loading factor (i.e. number of pairs per proton), and $R$ is the characteristic size of the corona. Introducing the normalized radius $\mathcal{R}\equiv R/R_s$, the coronal cold proton density is estimated as $n_{p,c} \simeq 10^9~{\rm cm}^{-3} \, \tau_{T,-1} \zeta_{e,0}^{-1} \mathcal{R}_{1.5}^{-1}$, where we used the notation $q_x = q/10^x$. 

Given that the produced neutrino has a typical energy $E_\nu \sim E_p/20$, we require $E_{p, \min} \sim 20$~TeV to reproduce the low-energy end of of the neutrino spectrum observed by IceCube. Because the spectrum of secondary particles produced in pp collisions follows closely the proton spectrum \citep{Kelner_2006}, we adopt the same slope for the parent proton distribution as the best-fit value for the spectral index of the neutrino spectrum determined by IceCube (i.e., $s_p=s_{\nu}=3.2$). We assume that the luminosity of the accelerated protons is comparable to the intrinsic X-ray luminosity of the corona, namely $L_p =10^{43.8}$~erg s$^{-1}$.  

Following \citep{2022ApJ...941L..17M} we introduce the parameter $\xi_{B}$, which is defined as the ratio of magnetic field energy density, $U_B = B^2/(8\pi)$, to the bolometric energy density of photons, $U_{ph} = L_{bol}/(4 \pi R^2 c)$. Here, $B$ represents the magnetic field strength, and the photon field is assumed to be distributed within a spherical source of radius $R$. Given these considerations, the magnetic field strength can be expressed as 
 $B \approx 1.4 \times 10^3$~G $\xi_{B}^{1/2} L_{bol,45}^{1/2} M_{7.3}^{-1}\mathcal{R}_{1.5}^{-1}$. Photons from neutral pion decay may pair produce on the disk-corona photon field. For our example, we adopt the spectral template for $L_{bol}=10^{45}$~erg s$^{-1}$ from \citet{Murase_2020} (see Fig.~2 therein). Pairs produced via $\gamma \gamma$ pair production and charged pion decays emit synchrotron radiation and inverse Compton scatter photons to high-energy, thus leading to the development of an electromagnetic cascade in the coronal region. Finally, all charged particles are assumed to escape from the corona on a timescale $t_{esc} = 100 R/c$, while photons and neutrinos escape on $R/c$.

\begin{figure}
\centering
\includegraphics[width=0.45\textwidth]{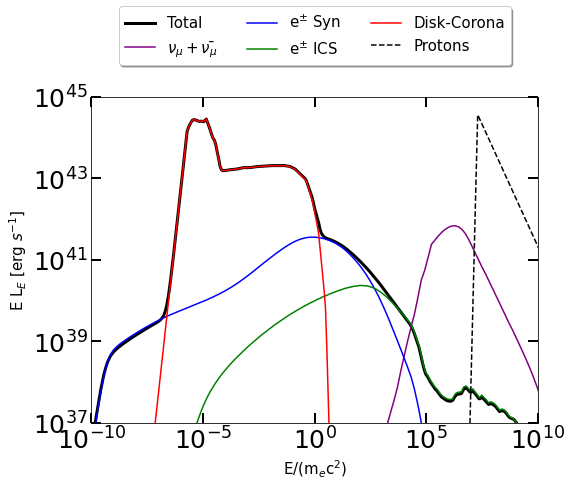} 
\caption{Spectral energy distribution of NGC 1068 is analyzed by considering an emitting region with $R=100 \, R_s$. The total photon spectrum originating from this region is shown by the solid black line. The solid red line represents the combined contribution of the corona and disk components (template adopted from \citet{Murase_2020}). Solid blue and green lines represent the synchrotron and ICS emission produced by the secondary particles, respectively. The neutrino spectrum is depicted by the purple line, and the black dashed line represents the injected luminosity of protons. }
\label{fig:NGC1068_SED}
\end{figure}

In Fig.~\ref{fig:NGC1068_SED} we present the broadband SED of our model for the case of an extended corona with radius $R=100~R_s$ and $\xi_B=1$. Neutral pions decay into $\gamma$-ray photons with a spectrum similar to the neutrino spectrum (solid magenta line). These energetic photons are attenuated by low-energy photons present (e.g. attenuation due to the disk thermal emission is evident by the dip at energies $E/(m_e c^2) \approx 10^6$), leading to the production of secondary pairs that emit synchrotron radiation (solid blue line) and inverse Compton scatter photons from the disk-corona system (solid green line). For reference, we also plot the differential proton luminosity that is injected in the corona (dashed black line).

\begin{figure}
\centering
\includegraphics[width=0.45\textwidth]{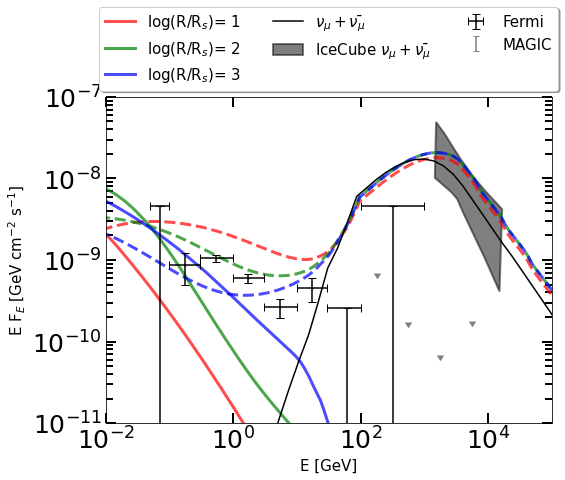} 
\caption{$\gamma$-ray and neutrino emission of NGC 1068. Black symbols and grey arrows indicate gamma-ray measurements by \fermi-LAT \citep{2020ApJS..247...33A} and upper limits by MAGIC \citep{2019ApJ...883..135A}, respectively. Colored solid (dashed) lines represent the $\gamma$-ray spectrum produced by the electromagnetic cascade inside the corona with (without) $\gamma\gamma$ absorption respectively, for different choices of radius and magnetic field (see legend and text for details). The solid dark line represents the muon and anti-muon neutrino spectrum, and the grey-shaded region covers all power-law neutrino fluxes that are consistent with the IceCube data at 95$\%$ confidence level \citep{2022Sci...378..538I}. }
\label{fig:NGC1068_neutrino}
\end{figure}

The high-energy photon and neutrino spectra of the model are compared to multi-messenger data of NGC 1068 in Fig. \ref{fig:NGC1068_neutrino} for $\xi_B=1$ and different values of the source radius, to illustrate the effects of $\gamma$-ray attenuation. For comparison, we also show the photon spectrum without accounting for $\gamma\gamma$ absorption, represented by the dashed colored lines. Our results reveal that the emission within the 0.1-10 GeV range is primarily generated through synchrotron radiation. In this particular model, the magnetic field strength is inversely proportional to the radius of the source.  As a result, stronger magnetic fields are expected in more compact regions (with smaller radii), pushing the synchrotron spectrum of secondaries to higher energies. As anticipated, we observe a greater degree of $\gamma\gamma$ absorption in more compact regions, but in all cases, the expected TeV emission from neutral pion decays is attenuated, hence all models are consistent with the MAGIC upper limits. We note that we have not included other potential sources of attenuation, such as the thermal emission from the dust torus. The model neutrino spectrum falls within the IceCube uncertainty band (95$\%$ confidence level) as expected. It is worth noting that the computed neutrino spectrum exhibits a spectral break at 0.1 TeV. This feature arises from the adoption of the $\delta-$function approximation for pion production at energies below this limit, as described in \citep{Kelner_2006} -- see also Appendix \ref{p-p_col}. Because of this assumption, the neutrino spectrum below 0.1 TeV is harder than the one presented in \citet{2022ApJ...941L..17M}, where the full neutrino spectrum produced by the minimum energy pions in the system was computed.

\section{Conclusions}\label{sec:conclusions}

In this work, we have presented our newly developed code, \code, which aims to describe the multi-messenger emission of non-thermal astrophysical sources. \code \ solves the kinetic equations using an implicit scheme assuming that the particles occupy a spherical region and they are homogeneous. The highlights of our work include the comprehensive treatment of various non-thermal processes, the efficient computational times achieved for different scenarios, the inclusion of energy-conserving schemes, and the adiabatic expansion of the source. These features make \code \  a valuable tool for studying non-thermal sources and their associated emission mechanisms.

One of the notable achievements of \code \  is the significant reduction in computation times compared to the ATHE$\nu$A code. We have demonstrated that \code \   can solve steady-state problems, involving leptons inside the emission region, in approximately $\sim 2-3$ seconds which is a factor of $30-100$ shorter than ATHE$\nu$A. For proton-proton steady-state scenarios, \code \  achieves computation times on the order of $\sim 20$ seconds, and for photohadronic steady-state scenarios, it takes approximately $\sim 10$ minutes (factor of 3 faster than ATHE$\nu$A). The improvement in the execution time can be attributed to the following factors. In comparison with \code, the time discretization in ATHE$\nu$A is not user-configurable, instead, it is predefined within the algorithm\footnote{ATHE$\nu$A utilizes a PDE solver from the Fortran NAG library that employs a variable-order, variable-time step method implementing the backward differentiation formulae}. It is established based on the most significant derivative encountered, considering all processes and grid points. As a result, if the initial conditions result in substantial derivatives, the time step becomes quite small, often extending the computation time required to reach a steady state. Typically, this results in computations taking a considerable amount of time, sometimes spanning the first few light-crossing times. On the contrary, the steady-state solutions are not sensitive to the choice of time step within the Chang \& Cooper scheme employed in \code; hence, larger time steps can be employed leading to faster computations for steady-state problems. Besides the different numerical schemes, there are discrepancies in the way certain processes are incorporated within each code. A more detailed explanation of the process implementation in \code \ can be found in Appendix \ref{appA}. Notable distinctions are the injection rates of the Bethe-Heitler and the photopion processes. In the ATHE$\nu$A code the Bethe-Heitler pair production rates and photopion production rates are based on tabulated Monte-Carlo results from \cite{1996APh.....4..253P} and \cite{2000CoPhC.124..290M}, respectively \cite[for the implementation in ATHE$\nu$A see][]{Mastichiadis_2005, DMPR12}.  In \code \ we use instead the analytical parametrizations of \cite{Kelner_2008} for the energy distributions of secondary particles produced from photopion interactions and simplify the calculation of pair distribution from Bethe-Heitler production as described in \ref{appA}. The improved efficiency of \code \  allows for more extensive simulations and understanding of the high-energy astrophysical sources.

The small computational time of \code \ enables the application of advanced data fitting techniques, such as Markov Chain Monte Carlo (MCMC) methods. Using {\tt emcee} we have demonstrated the capability of \code \  to fit the SED of a blazar, using two different scenarios and taking into account the observations in the X-rays and $\gamma$-rays. By employing MCMC techniques, we gain a deeper understanding of the underlying physical parameters and their associated uncertainties, enabling us to effectively reproduce the observed spectrum. Furthermore, our analysis revealed that the spread in fluxes across the electromagnetic spectrum can be larger than the systematic differences of $10\%$ between radiative transfer codes, including the ATHE$\nu$A code. Additionally, our results emphasize the importance of dense sampling of the SED through multiwavelength observing campaigns to capture the full complexity of non-thermal sources.

Other significant aspects of the code are the inclusion of energy-conserving treatment between the particles and the photons and the consideration of photons that are produced from the secondary particles as targets for particle-photon interactions. This latter feature allows the study of optically thick environments where feedback effects are important for driving oscillatory (limit-cycle) behavior found in hadronic supercriticalities \citep{PM12, Mastichiadis_2005, 2020MNRAS.495.2458M}. By accurately accounting for these interactions, \code \ may provide a more comprehensive description of the photon spectra escaping such environments.

 Despite the advancements achieved in \code \ there are still challenges to be addressed. The computational time needed for photohadronic scenarios currently prohibits the MCMC fitting applied to high-energy astrophysical sources. Furthermore, small changes in the physical parameters may lead to different spectral characteristics due to the interplay of the physical processes. To overcome these challenges, one potential way to move forward is to explore the use of neural networks as an alternative to numerical codes. Neural networks have gained traction in various fields, such as astrophysics \citep{2017Natur.548..555H}, space weather forecasting \citep{Camporeale_2018}, and biology \citep{2019NatCo..10.4354W}. Implementing neural networks in the analysis of high-energy astrophysical emission could provide a more efficient approach to fitting big data sets by reducing significantly the computation time while providing an accuracy comparable to the uncertainties of emission models.

\begin{acknowledgements}
We thank the anonymous referee for his/her useful feedback. We thank Kohta Murase for insightful discussions about NGC 1068. We also thank Stella Boula and Paolo Padovani for useful comments on the manuscript. S.I.S. and M.P. acknowledge support from the Hellenic Foundation for Research and Innovation (H.F.R.I.) under the ``2nd call for H.F.R.I. Research Projects to support Faculty members and Researchers'' through the project UNTRAPHOB (Project ID 3013). GV acknowledges support by H.F.R.I. through the project ASTRAPE (Project ID 7802).
\end{acknowledgements}

%
%

\bibliographystyle{aa} 
\bibliography{ref.bib}

\appendix

\section{Radiative processes}\label{appA}

\subsection*{Adiabatic losses}
Particles inside an expanding spherical region of radius $R$ will undergo energy losses as they do work. The energy losses of particle species $i$ are related to the expansion rate and are proportional to the energy of the particle and are given by \citep{2011hea..book.....L}:

\begin{equation}
\frac{d\epsilon_{i}}{dt}\Bigg|_{ad} = -\frac{1}{R}\frac{dR}{dt}\epsilon_i,
\label{eq:dg_dt_ad}
\end{equation}
where $\epsilon_i=\gamma_i$ is the particle Lorentz factor of particle species $i$ with rest mass energy $m_i c^2$ (or frequency $\nu$ for photons and neutrinos). Therefore, the loss term that appears in Eqs.~\ref{eq:gen_kin_eq_e}, \ref{eq:gen_kin_eq_p} and \ref{eq:gen_kin_eq_nu} can be expressed as follows

\begin{equation}
\mathcal{L}_{i}^{ad} \equiv \frac{dN_i}{dtd\epsilon_i} = \frac{\partial}{\partial \epsilon_i}\Bigg(\frac{d\epsilon_{i}}{dt}\Bigg |_{ad} N_i\Bigg),
\label{eq:L_ad_i}
\end{equation}

\subsection*{Synchrotron emission}
The synchrotron emissivity of a relativistic particle species $i$ with rest mass $m_i$,  charge $e$, and differential number density $n_i(\gamma_i) \equiv N_i(\gamma_i)/V$ (where $V$ is the volume of the emitting region) is given by \citep{boettcher2012relativistic}:

\begin{equation}
\begin{split}
j^{syn}_{\nu,i} \equiv \frac{dN}{d\Omega dVdtd\nu} = \frac{c\sigma{_T}U_{B}}{3\pi h\Gamma(4/3)}\bigg(\frac{m_e}{m_i}\bigg)^2\nu^{-2/3}\times \\ 
\int^{\infty}_1d\gamma_i  \, n_{i}(\gamma_i) \, \gamma^2\frac{e^{-\nu/\nu_{c}}}{\nu_{c}},
\label{eq:Syn_emis}
\end{split}
\end{equation}
where $m_i$ is the mass of the charged particle, $\Gamma(x)$ is the Gamma function, $U_B=B^2/8\pi$ is the magnetic energy density,  and $\nu_{c}$ is the critical synchrotron frequency

\begin{equation}
\nu_{c} = \frac{3eB}{4\pi m_i c }\gamma^2.
\label{eq:crit_freq}
\end{equation}
Given the above, we can now define the synchrotron source and loss terms that appear in Eqs.~\ref{eq:gen_kin_eq_e}, \ref{eq:gen_kin_eq_p}, and \ref{eq:gen_kin_eq_nu} as

\begin{equation}
Q_{\gamma}^{i,syn} \equiv \frac{dN}{dtd\nu} = \int j^{syn}_{\nu,i} dV d\Omega,
\label{eq:Q_syn_gamma}
\end{equation}

\begin{equation}
\mathcal{L}_{i}^{syn} \equiv \frac{dN_i}{dtd\gamma_i} = \frac{\partial}{\partial \gamma_i}\Bigg(\frac{d\gamma_{i}}{dt}\Bigg |_{syn} N_i\Bigg),
\label{eq:L_syn_i}
\end{equation}
where $d\gamma_{i}/dt$ are the energy losses of the particle species $i$ due to synchrotron radiation

\begin{equation}
\frac{d\gamma_{i}}{dt}\Bigg |_{syn} = -\frac{4\sigma_{T}U_{B}}{3m_ec}\gamma_i^2\Bigg(\frac{m_e}{m_i}\Bigg)^3.
\label{eq:dg_dt_syn}
\end{equation}

\subsection*{Synchrotron self-absorption}
Low-energy charged particles in the plasma can re-absorb some of the emitted synchrotron radiation. This process is known as synchrotron self-absorption. The synchrotron self-absorption coefficient (often denoted as $\alpha_{ssa}(\nu)$) is a measure of how much of the radiation produced by a synchrotron source is absorbed by the emitting plasma itself, rather than being observed by an external observer. The coefficient for a charged particle with mass $m{_i}$ is given by \citep{dermer2009high}:

\begin{equation}
a_{\nu ,ssa}=-\frac{1}{8\pi m_i\nu^2}\int_{1}^{\infty}d\log\gamma_i
\Bigg[\frac{dn_i(\gamma_i)}{d\log\gamma_i}-2n_i(\gamma_i)\Bigg]P_i(\nu,\gamma_i),
\label{eq:ssa_coef}
\end{equation}
where $P_i$ is the synchrotron power spectrum emitted by a single particle (see Eq. 2.15 and Eq. 2.18 in \citet{ghisellini2008beaming}) assuming only head-on collisions. We can then define the optical depth as follows
\begin{equation}
\tau_{\nu, ssa}=a_{\nu ,ssa} R.
\label{eq:opt_depth_ssa}
\end{equation}
The photon spectrum is strongly absorbed below a frequency at which the optical depth is above unity ($\tau_{\nu, ssa}\geq 1$). We then introduce the synchrotron self-absorption loss term in Eq. \ref{eq:gen_kin_eq_nu} as follows

\begin{equation}
\mathcal{L}_{\gamma}^{ssa} \equiv \frac{dN}{dtd\nu} = N_{\gamma}a_{\nu ,ssa}c.
\label{eq:L_ssa_gamma}
\end{equation}

\subsection*{Inverse Compton scattering}
Charged particles can also up-scatter photons via Inverse Compton (IC) scattering. In this work, we consider only IC scattering by electrons (and positrons), since proton IC scattering is negligible for most astrophysical sources \citep{Kelner_2008} \citep[see, however,][]{AM3}.

In the general case, where we use the exact Klein-Nishina formula for the cross-section, the scattered photon spectrum per electron can be expressed using the frequency of the scattered photon in units of the initial electron energy ($ \nu_1=\gamma_e m_e c^2E_1/h$). Assuming that the photons undergo IC scattering by an electron population with differential number density $n_e(\gamma_e)$, then the emissivity of IC scattering is given by \citep{RevModPhys.42.237}:

\begin{equation}\label{eq:IC_emis}  
\begin{split}
j_{\nu_1} \equiv \frac{dN}{dt d\nu_1 dV}=\sigma_{T}c\int^{\infty}_1 n_e(\gamma_e) \int_0^{\infty} \frac{ n_{\gamma}(\nu)}{4\gamma_e^2\nu}[2q ln(q)+\\ (1+2q)(1-q)+\frac{1}{2}\frac{(\Gamma_{\epsilon}q)^2}{1+\Gamma_{\epsilon}q}(1-q)]d\nu\ d\gamma_e ,
\end{split}
\end{equation}
where $n_e(\gamma_e)\equiv dN_e/dVd\gamma_e$, $n_{\gamma}(\nu)\equiv dN_{\gamma}/dVd\nu$ , $\Gamma_{\epsilon}=4h\nu\gamma_e/m_ec^2$, and  $q=E_1/\Gamma_{\epsilon}(1-E_1)$.

We, therefore, introduce the inverse Compton scattering source term in Eq. \ref{eq:gen_kin_eq_nu} as follows

\begin{equation}
Q_{\gamma}^{ics} \equiv \frac{dN}{dtd\nu_1} = \int j_{\nu_1} dV .
\label{eq:Q_IC_gamma}
\end{equation}

We account only for energy losses that are occurring in the Thomson regime. The Lorentz factor of electrons undergoing IC scattering in this regime can be expressed as follows

\begin{equation}
\frac{d\gamma_e}{dt}\Bigg |_{ics} = -\frac{4}{3}\sigma_T c U_{ph}(\gamma_e)\gamma_e^2,
\label{eq:IC_losses}
\end{equation}
where $U_{ph}(\gamma_e)$ is given by \citep{1995A&A...295..613M}:

\begin{equation}
U_{ph}(\gamma_e) = \frac{h}{m_ec^2}\int_0^{\frac{3m_ec^2}{4\gamma_eh}}d\nu' \nu' n_{\gamma}(\nu').
\label{eq:U_ph_Thomson}
\end{equation}

Therefore, the loss term that appears in Eq. \ref{eq:gen_kin_eq_e} can be written as follows

\begin{equation}
\mathcal{L}_{e}^{ics} \equiv \frac{dN_e}{dtd\gamma_e} = \frac{\partial}{\partial \gamma_e}\Bigg(\frac{d\gamma_{e}}{dt}\Bigg |_{ics} N_e\Bigg).
\label{eq:L_ics_e}
\end{equation}

\subsection*{Photon-photon pair production} 
A process that acts as a sink of high-energy photons, and as a source term for relativistic electrons is photon-photon pair creation. We introduce the absorption coefficient for a given high energy photon with frequency $\nu_1$ as follows

\begin{equation}
a_{\gamma \gamma}=-\int_{0}^{\infty}d \epsilon \ n_{\gamma}(\epsilon )\sigma_{\gamma \gamma}(y),
\label{eq:gg_coef}
\end{equation}

where $\sigma_{\gamma \gamma}$ is the cross-section of the process, $y=\epsilon \epsilon_1$ with $\epsilon$ and $\epsilon_1$ being the target and the $\gamma$-ray photon energies respectively normalized to the electron's rest mass energy. We use an approximation for the cross section from \citep{1990MNRAS.245..453C}: 

\begin{equation}
\sigma_{\gamma \gamma}(y) \simeq 0.652\sigma_T\frac{y^2-1}{y^3}\ln(y){y}\Theta(y-1),
\label{eq:gg_cross_section}
\end{equation}
where $\Theta(y)$ is the Heavyside function. We, therefore, introduce the $\gamma \gamma$ absorption loss term in Eq.~\ref{eq:gen_kin_eq_nu} that reads

\begin{equation}
\mathcal{L}_{\gamma}^{\gamma \gamma} \equiv \frac{dN}{dtd\nu_1} = N_{\gamma}a_{\gamma \gamma}c.
\label{eq:L_gg_gamma}
\end{equation}

The volumetric production rate of relativistic electrons (or positrons) of Lorentz factor $\gamma_e$ is given by \citep{1995A&A...295..613M}:

\begin{equation}\label{eq:gg_emis}
\begin{split}
j_{\gamma_e}^{\gamma \gamma}
\equiv\frac{dN}{d\gamma_e dt dV}=4\sigma_T c n_{\gamma}(2\gamma_e,t)\frac{m_e c^2}{h}\times \\ \int_{\frac{m_e c^2}{2\gamma_e h}}^{\infty}d\nu_1' n_{\gamma}(\nu_1',t)R_{\gamma \gamma}\bigg(\frac{2\gamma_e h \nu_1'}{m_e c^2}\bigg),
\end{split}
\end{equation}
where $R_{\gamma \gamma}$ is an approximation to reaction rate of the process \citep{1995A&A...295..613M}. In the previous equation, all angle effects are neglected and we have assumed that the emerging pair has energy equal to that of the absorbed $\gamma$-ray. This leads to the following source term in Eq.~\ref{eq:gen_kin_eq_e}

\begin{equation}
Q_{e}^{\gamma \gamma} \equiv \frac{dN}{dtd\gamma_e} = \int j_{\gamma_e}^{\gamma \gamma} dV,
\label{eq:Q_ee_gg}
\end{equation}

\subsection*{Bethe-Heitler pair production}
\begin{figure}
\centering
\includegraphics[width=8.5cm]{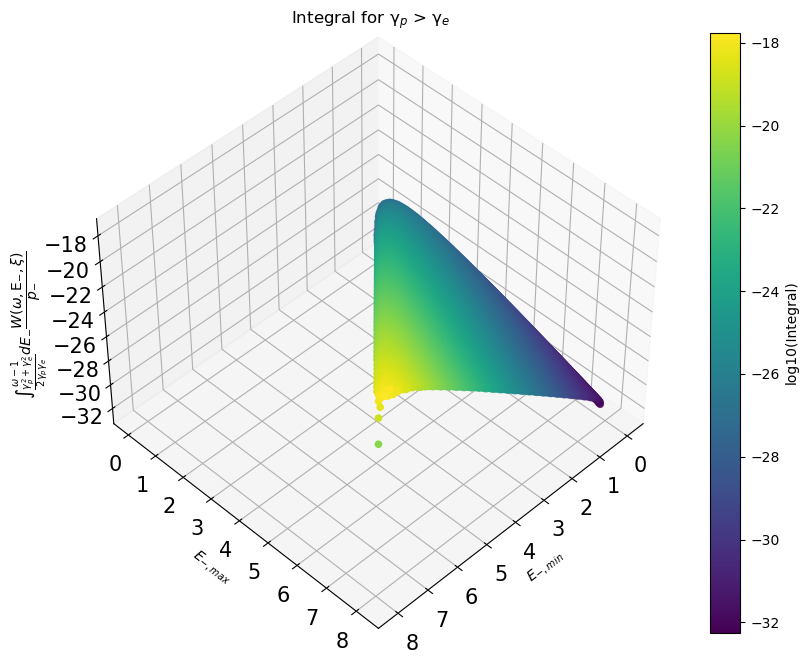}\\
\includegraphics[width=8.5cm]{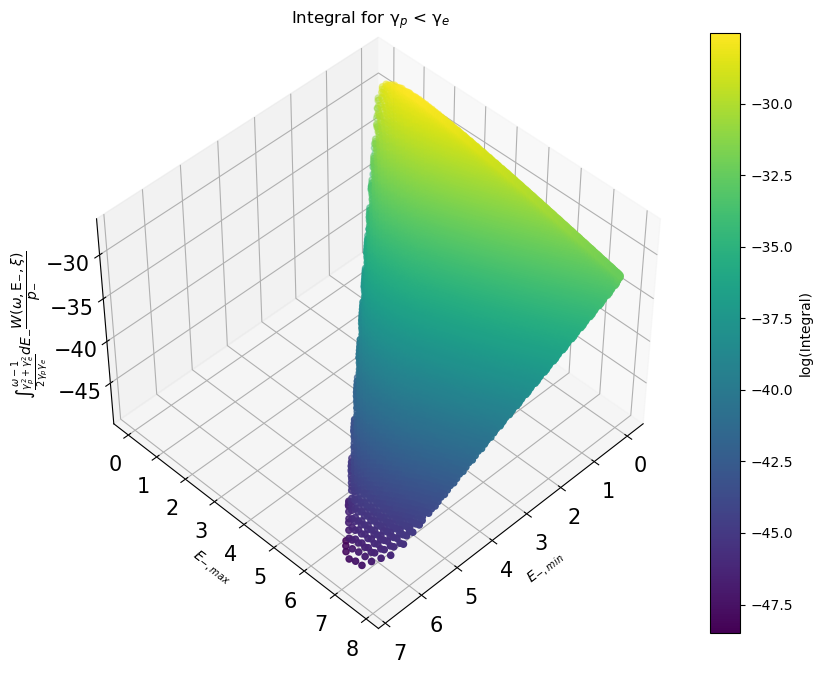}\\
\caption{Third integral of Eq. \ref{eq:gg_emis} for several values of the integration limits here denoted as $E_{-,min}$ and $E_{-,max}$. The top panel shows the integral's value for $\gamma_p>\gamma_e$, while the bottom panel shows the integral's values for $\gamma_p<\gamma_e$.} 
\label{fig:Int_BH}       
\end{figure}

Bethe-Heitler pair production refers to the process of creating an electron-positron pair by a relativistic proton interacting with a low-energy photon. For ultra-relativistic protons $\gamma_p \gg 1$, the pair spectrum is given by \citep{Kelner_2008}:
\begin{equation}
\frac{dN}{d\gamma_e}=\int_{\frac{(\gamma_{p}+\gamma_e)^2}{4\gamma_{p}^2\gamma_e}} ^{\frac{m_p}{\gamma_{p}m_e}} d\epsilon \frac{n_{\gamma}(\epsilon)}{\epsilon^2}\int_{\frac{(\gamma_{p}+\gamma_e)^2}{2\gamma_{p}\gamma_e}}^{2\gamma_{p}\epsilon}d\omega\ \omega\int_{\frac{(\gamma_{p}^2+\gamma_e^2)}{2\gamma_{p}\gamma_e}} ^{\omega-1} dE_{-} \frac{W(\omega ,E_{-} ,\xi )}{p_{-}},
\label{eq:BH_spec}
\end{equation}
where $W(\omega ,E_{-} ,\xi )$ is the cross-section \citep{PhysRevD.1.1596} and all energies above are normalized to the electron rest mass energy. The last integral in Eq.~\ref{eq:BH_spec}, involving the cross-section, can be computationally challenging and can increase the computation time of a simulation significantly. To overcome this problem, we calculate the integral for various combinations of $\gamma_e$, $\gamma_p$, and $\omega$, from a large sample, given the parameters of each simulation, and store them in an array. In Fig. \ref{fig:Int_BH}, we illustrate the third integral from Eq. \ref{eq:BH_spec} for several values of $E_{-,min} \equiv 
 \frac{(\gamma^2_p+\gamma^2_e)}{2\gamma_p \gamma_e}$ and $E_{-,max} \equiv \omega-1$. The upper panel of Fig. \ref{fig:Int_BH} depicts the integral's value when $\gamma_p>\gamma_e$, whereas the lower panel represents the scenario with the reverse condition. Notably, we observe a division of the integral into two distinct regions at the plane where $\gamma_p=\gamma_e$. By using interpolation, the desired integral values can be obtained efficiently, thus reducing the computational time by a factor of $\sim 20-30$. 

In order to calculate the volumetric production rate of pairs coming from a proton distribution $n_p(\gamma_p)\equiv N_p(\gamma_p)/V$ we use

\begin{equation}
j_{\gamma_e}^{bh} \equiv \frac{dN}{d\gamma_e dt dV}= 2c\int_{max(1,\gamma_p>\gamma_e m_e/m_p)}^{\infty}d\gamma_p n_p(\gamma_p)\frac{dN}{d\gamma_e},
\label{eq:BH_emis}
\end{equation}
Therefore for the source term of relativistic electrons in Eq.~ \ref{eq:gen_kin_eq_e} we use

\begin{equation}
Q_{e}^{bh} \equiv \frac{dN}{dtd\gamma_e} = \int j_{\gamma_e} dV.
\label{eq:Q_ee_bh}
\end{equation}
The energy loss rate of a high energy proton with Lorentz factor $\gamma_p$ is given by \citep{PhysRevD.1.1596}:

\begin{equation}
\frac{d\gamma_p}{dt}\Bigg|_{bh} = -\frac{3\sigma_T c m_e}{8\pi\alpha_f m_p}\int_2^{\infty}n_{\gamma}\bigg(\frac{\xi m_e c^2}{2\gamma_p}\bigg)\frac{\phi(\xi)}{\xi^2}d\xi,
\label{eq:BH_losses}
\end{equation}
where $\alpha_f$ is the fine structure constant and the function $\phi(\xi)$ is adopted from Fig. 2 of \citep{PhysRevD.1.1596}.
The energy loss term in Eq.~\ref{eq:gen_kin_eq_p} can be then written as

\begin{equation}
\mathcal{L}_{p}^{bh} \equiv \frac{dN_p}{dtd\gamma_p} = \frac{\partial}{\partial \gamma_p}\Bigg(\frac{d\gamma_{p}}{dt} \Bigg|_{bh} N_p\Bigg).
\label{eq:L_bh_p}
\end{equation}

\subsection*{Photomeson production process}
Photomeson interactions refer to the process by which low-energy photons interact with high-energy protons, resulting in the production of mainly $\pi^0$ and $\pi^{\pm}$ mesons. The produced pions will decay in $\gamma$-rays and leptons. The production rate of a given type of particle with energy $E_l$, where $l$ can be $\gamma,e^+,e^-,\nu_{\mu},\bar{\nu}_{\mu},\nu_{e}$ and $\bar{\nu}_{e}$, from a proton distribution $n_{p}(E_p) \equiv N_p(E_p)/V$ is given by \citep{Kelner_2008}:

\begin{equation}
\frac{dN_l}{dtdVdE_l}=\int n_p(E_p)n_{ph}(\epsilon)\Phi_{l}(\eta,x)\frac{dE_p}{E_p}d\epsilon,
\label{eq:photomeson_emis}
\end{equation}
where $E_p=\gamma_p m_p c^2$,  $\epsilon$ is the photon energy normalized to $m_e c^2$, $x$ is the ratio between the energy of the produced particle and the proton's energy and $\Phi_{l}(\eta,x)$ is the energy distribution of the species $l$ -- see Eq.~(31) and Tables I-III in \cite{Kelner_2008}.

Therefore for the source term of relativistic particles with energy $E_l$ in Equations \ref{eq:gen_kin_eq_e}, \ref{eq:gen_kin_eq_nu}, and \ref{eq:gen_kin_eq_g} we use

\begin{equation}
Q_{l}^{p\gamma,\pi} \equiv \frac{dN}{dtdE_l} = \int \frac{dN_l}{dtdVdE_l} dV.
\label{eq:Q_pg_l}
\end{equation}

The energy loss rate of a high energy proton with Lorentz factor $\gamma_p$ are given by \citep{1990ApJ...362...38B}:

\begin{equation}
\frac{d\gamma_p}{dt}\Bigg|_{p\gamma,\pi}  = -\frac{c}{2\gamma_p}\int_{\bar{\epsilon}_{th}/h}^{\infty}d\bar{\nu}\sigma_{p \pi}(\bar{\nu})k_p(\bar{\nu})\bar{\nu}\int_{\bar{\nu}/2\gamma_p}^{\infty}d\nu'\frac{n'_{\gamma}(\nu')}{\nu'^2},
\label{eq:pg_losses}
\end{equation}
where $n'_{\gamma}(\nu')\equiv dN_{\gamma}/dVd\nu'$ denotes the density of photons in the emission region's frame of reference while all bared quantities are measured in the proton's rest frame. The energy-dependent cross-section and the inelasticity of the process are represented as $\sigma_{p\pi}$ and $k_p$ and are adopted from \citep{2019JCAP...11..007M} and \citep{PhysRevLett.21.1016} respectively. 

The energy loss term in Eq.~\ref{eq:gen_kin_eq_p} can then be written as

\begin{equation}
\mathcal{L}_{p}^{p\gamma,\pi} \equiv \frac{dN_p}{dtd\gamma_p} = \frac{\partial}{\partial \gamma_p}\Bigg(\frac{d\gamma_{p}}{dt}\Bigg|_{p\gamma,\pi}N_p\Bigg).
\label{eq:L_pg_p}
\end{equation}

\subsection*{Proton-proton (pp) inelastic collisions}\label{p-p_col}
High-energy protons may interact with non-relativistic (cold) protons in dense astrophysical environments. Inelastic proton-proton collisions lead to the production of secondary particles, $\gamma$-rays, pairs, and neutrinos.  The production rate of a given type of particle with energy $E_l$, where $l$ can be $\gamma,e^+,e^-,\nu_{\mu},\bar{\nu}_{\mu},\nu_{e}$ and $\bar{\nu}_{e}$, from a proton distribution $n_{p}(E_p) \equiv N_p(E_p)/V$ is given by \citep{2006PhRvD..74c4018K}:

\begin{equation}
\frac{dN_l}{dtdVdE_l}= c\ n_g \int_0 ^1 \sigma_{pp, inel}(E_{l}/x) n_p(E_{\gamma}/x) F_{l}(x, E_{l}/x) \frac{dx}{x},
\label{eq:pp_emis}
\end{equation}
where $E_p=\gamma_p m_pc^2$, $x= E_{l}/E_p$, $n_{g}$ is the number density of cold protons in the ambient gas, and $\sigma_{pp, inel}$ is the cross-section of inelastic pp interactions. The function $F_{l}(x, E_{l}/x)$ is related to the energy distribution of particle species $l$. We use the empirical functions as described in \citep{2006PhRvD..74c4018K} for $E_p > 0.1$~TeV, and $x_l = E_l/E_p \geq 10^{-3}$, while for values $E_p < 0.1$~TeV the spectra of $\gamma$-ray photons and other secondaries are computed to lower energies using the $\delta$-function approximation for pion production as described in the same manuscript.

Therefore, the source term of relativistic particles with energy $E_l$ in Eqs.~\ref{eq:gen_kin_eq_e}, \ref{eq:gen_kin_eq_nu} and \ref{eq:gen_kin_eq_g} is given by,

\begin{equation}
Q_{l}^{pp} \equiv \frac{dN}{dtdE_l} = \int \frac{dN_l}{dtdVdE_l} dV.
\label{eq:Q_pp_l}
\end{equation}

The energy loss rate of a high energy proton with Lorentz factor $\gamma_p \gg 1$ is given by \citep[see Eq.~4.11 in][]{1994A&A...286..983M}:

\begin{equation}
\frac{d\gamma_p}{dt}\Bigg|_{pp} = -\frac{0.65}{m_pc}\ n_g\sigma_{pp, inel}(E_p)\Theta(E_p-E_{p,th}),
\label{eq:pp_losses}
\end{equation}
where $E_{p,th}= \gamma_{p, th} m_p c^2$ and $\gamma_{p, th} = 1 + (m_{\pi}/ m_p)(1/2 + m_\pi/(2 m_p))$. 

The energy loss term in Eq.~\ref{eq:gen_kin_eq_p} can be written as

\begin{equation}
\mathcal{L}_{p}^{pp} \equiv \frac{dN_p}{dtd\gamma_p} = \frac{\partial}{\partial \gamma_p}\Bigg(\frac{d\gamma_{p}}{dt}\Bigg|_{pp} N_p\Bigg).
\label{eq:L_pp_p}
\end{equation}

\section{Kinetic Equations:  number density versus particle number}\label{appB}
In this section, we examine the results obtained by numerically solving the kinetic equation when this is expressed with respect to the particle number density or to the particle number. As the first example, we demonstrate the numerical solution of Test 3 using the number density in the kinetic equation instead of the number of particles. We can discretize Eq.~\ref{eq:gen_kin_eq_e} into time $t_i$ and Lorentz factor $\gamma_j$ for the two cases. When the kinetic equation is written in terms of the particle number, the discretization reads

\begin{equation}
V_{1,j}N_{e,j-1}^{i+1}+V_{2,j}N_{e,j}^{i+1}+V_{3,j}N_{e,j+1}^{i+1}=N_{e,j}^i
\label{eq:k_e_N_e},
\end{equation}
where the coefficients are
\begin{equation}
V_{1,j} = 0,
\label{eq:V1_coef_N_e}
\end{equation}
\begin{equation}
V_{2,j} = 1+\frac{\Delta t}{\Delta \gamma_j}\frac{V_{exp}}{R^{i+1}}\gamma_{j},
\label{eq:V2_coef_N_e}
\end{equation}
\begin{equation}
V_{3,j} = -\frac{\Delta t}{\Delta \gamma_j}\frac{V_{exp}}{R^{i+1}}\gamma_{j+1}.
\label{eq:V3_coef_N_e}
\end{equation}

On the other hand, when we discretize the equation written in terms of the number density, we find

\begin{equation}
V_{1,j}n_{e,j-1}^{i+1}+V_{2,j}n_{e,j}^{i+1}+V_{3,j}n_{e,j+1}^{i+1}=n_{e,j}^i
\label{eq:k_e_n_e},
\end{equation}
where the coefficients are

\begin{equation}
V_{1,j} = 0,
\label{eq:V1_coef_n_e}
\end{equation}
\begin{equation}
V_{2,j} = 1+3\Delta t\frac{V_{exp}}{R^{i+1}}+\frac{\Delta t}{\Delta \gamma_j}\frac{V_{exp}}{R^{i+1}}\gamma_{j},
\label{eq:V2_coef_n_e}
\end{equation}
\begin{equation}
V_{3,j} = -\frac{\Delta t}{\Delta \gamma_j}\frac{V_{exp}}{R^{i+1}}\gamma_{j+1}.
\label{eq:V3_coef_n_e}
\end{equation}

\begin{figure}
\centering
\includegraphics[width=0.45\textwidth]{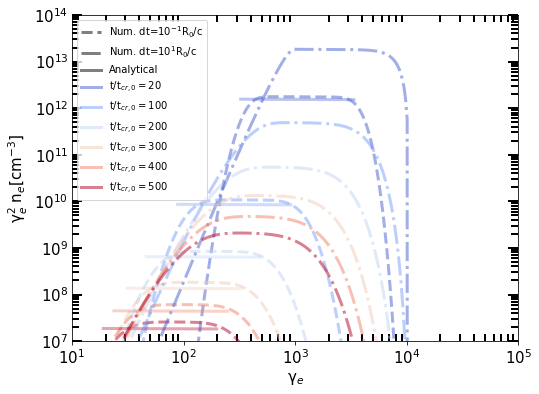} 
\caption{Comparison between numerical solutions (represented by dashed and dashed-dotted lines) and analytical solutions (represented by solid lines) using the same parameters as Test 3, using the number density in the kinetic equation. The numerical results were obtained with varying time steps, which are indicated in the legend.}
\label{fig:Test_3_kin_eq_n_e}
\end{figure}

In Figure \ref{fig:Test_3_kin_eq_n_e}, we depict the numerical solution obtained after solving Eq.~(\ref{eq:k_e_n_e}) with two choices of the time step (represented by dotted and dashed-dotted lines). The only disparity between the coefficients in Eqs.~(\ref{eq:k_e_N_e}) and (\ref{eq:k_e_n_e}) lies in the coefficient $V_{2,j}$. When solving the kinetic equation utilizing the number density, an extra term emerges from the time derivative of the volume, specifically $3\Delta t\frac{V_{exp}}{R^{i+1}}$. 
We observe that the numerical solution diverges by almost an order of magnitude in normalization from the analytical one when 
a large time step is chosen -- the difference is smaller when the kinetic equation with respect to the number is solved (compare to Fig.~\ref{fig:Test3_N_el}). To ensure a better agreement with the analytical solution, the time step has to satisfy $\Delta t < \frac{R_0}{3V_{exp}}$.

As a second example, we consider an expanding spherical source containing particles that undergo only adiabatic losses. We assume a constant injection of particles (e.g. electrons) with a power-law distribution of slope $s_e$ between $\gamma_{\min}$ and $\gamma_{\max}$. The source is expanding with velocity $V_{exp}=0.1c$. This problem has no steady-state solution and is also ideal for examining the performance of a numerical solver in time-dependent scenarios.

The analytical solution to the kinetic equation is given by the following expression
\begin{equation}
N_e(\gamma,t) = K\frac{1+\beta_0t}{\beta_0 s_e}[1-(1+\beta_0t)^{-s_e}]\gamma^{-s_e}, \gamma_{\min}\leq\gamma\leq\gamma_{\max},
\label{eq:k_e_analytical_sol}
\end{equation}
where $\beta_0\equiv V_{exp}/R_0$ and $K$ is a normalization constant. 

The discretize Eq.~\ref{eq:gen_kin_eq_e} into time $t_i$ and Lorentz factor $\gamma_j$ in terms of the particle number

\begin{equation}
V_{1,j}N_{e,j-1}^{i+1}+V_{2,j}N_{e,j}^{i+1}+V_{3,j}N_{e,j+1}^{i+1}=N_{e,j}^i+Q_{j}^i\Delta t,
\label{eq:k_e_N_e_inj}
\end{equation}
where the coefficients are given by Eqs.~\ref{eq:V1_coef_N_e}, \ref{eq:V2_coef_N_e} and \ref{eq:V3_coef_N_e}

On the other hand, when we discretize the equation written in terms of the number density, we find

\begin{equation}
V_{1,j}n_{e,j-1}^{i+1}+V_{2,j}n_{e,j}^{i+1}+V_{3,j}n_{e,j+1}^{i+1}=n_{e,j}^i+Q_{j}^i\Delta t
\label{eq:k_e_n_e_inj},
\end{equation}
where the coefficients are given by Eqs.~\ref{eq:V1_coef_n_e}, \ref{eq:V2_coef_n_e} and \ref{eq:V3_coef_n_e}

\begin{figure}
\centering
\includegraphics[width=0.45\textwidth]{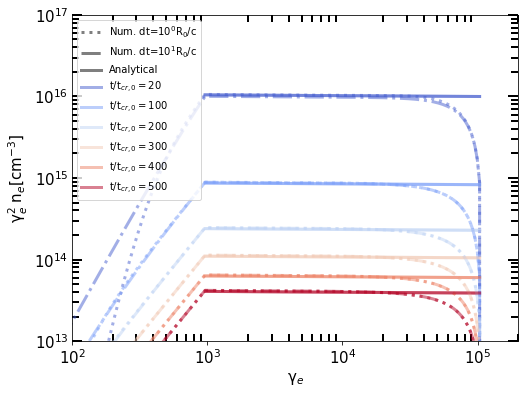} 
\caption{Comparison of numerical (dashed and dotted lines) resulted by solving Eq.~\ref{eq:k_e_N_e} and analytical (solid lines) solutions for adiabatic cooling electrons in an expanding blob with constant injection. Numerical solutions for two choices of the time step are shown as indicated in the legend. The evolution of the particle distribution in time is indicated by the different colors.}
\label{fig:kin_eq_N_e}
\end{figure}

\begin{figure}
\centering
\includegraphics[width=0.45\textwidth]{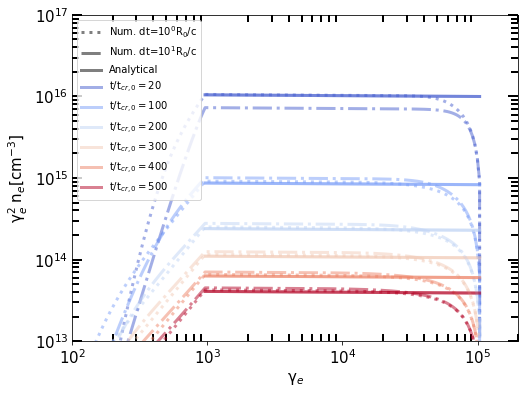} 
\caption{Comparison of numerical (dashed and dotted lines) resulted by solving eq. \ref{eq:k_e_n_e} and analytical (solid lines) solutions for adiabatic cooling electrons in an expanding blob with constant injection. Each numerical result was calculated using a different time step as indicated in the legend.}
\label{fig:kin_eq_n_e}
\end{figure}

Again the only difference between the coefficients in the equations above lies in the coefficient $V_{2,j}$. When solving the kinetic equation using the number density, an additional term arises from the time derivative of the volume $3\Delta t\frac{V_{exp}}{R^{i+1}}$. In figures \ref{fig:kin_eq_N_e} and \ref{fig:kin_eq_n_e}, we display both the numerical and analytical solutions for the two cases, considering two different time steps for each scenario. Our findings reveal that when using the particle number approach, the numerical and analytical solutions exhibit the same behavior for both choices of $\Delta t$. However, when solving the kinetic equation using the number density of particles, we observe a discrepancy between the analytical and numerical solutions for larger time steps. It is important to note that this discrepancy arises when the chosen time step $\Delta t>R_0/(3V_{exp})$. Consequently, we choose to solve the kinetic equation using the particle number rather than the number density, as it yields more accurate numerical solutions without the need for very small time steps.

\section{Corner plots}\label{section:C_P}
\begin{figure*}
\centering
\includegraphics[width=0.99\textwidth]{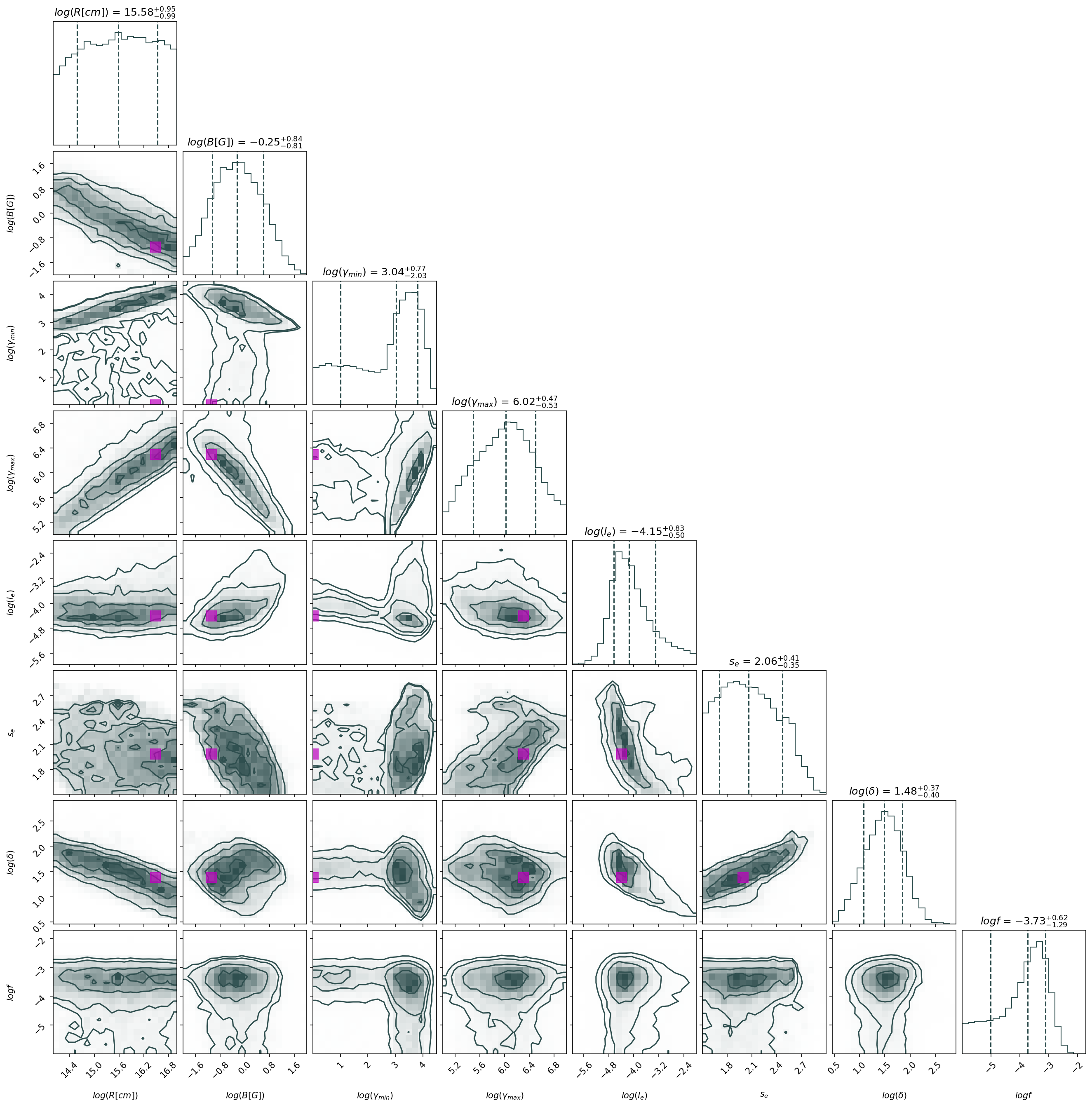} 
\caption{Corner plot showing the posterior distributions of the parameters of the SSC model of \hsp. Dashed lines in the histograms indicate the median value of each parameter and the 68 \% range of values. The parameter values used in model D from \citet{Petro2020} are overplotted (magenta square) for comparison.}
\label{fig:corner-ssc}
\end{figure*}

\begin{figure*}
\centering
\includegraphics[width=0.99\textwidth]{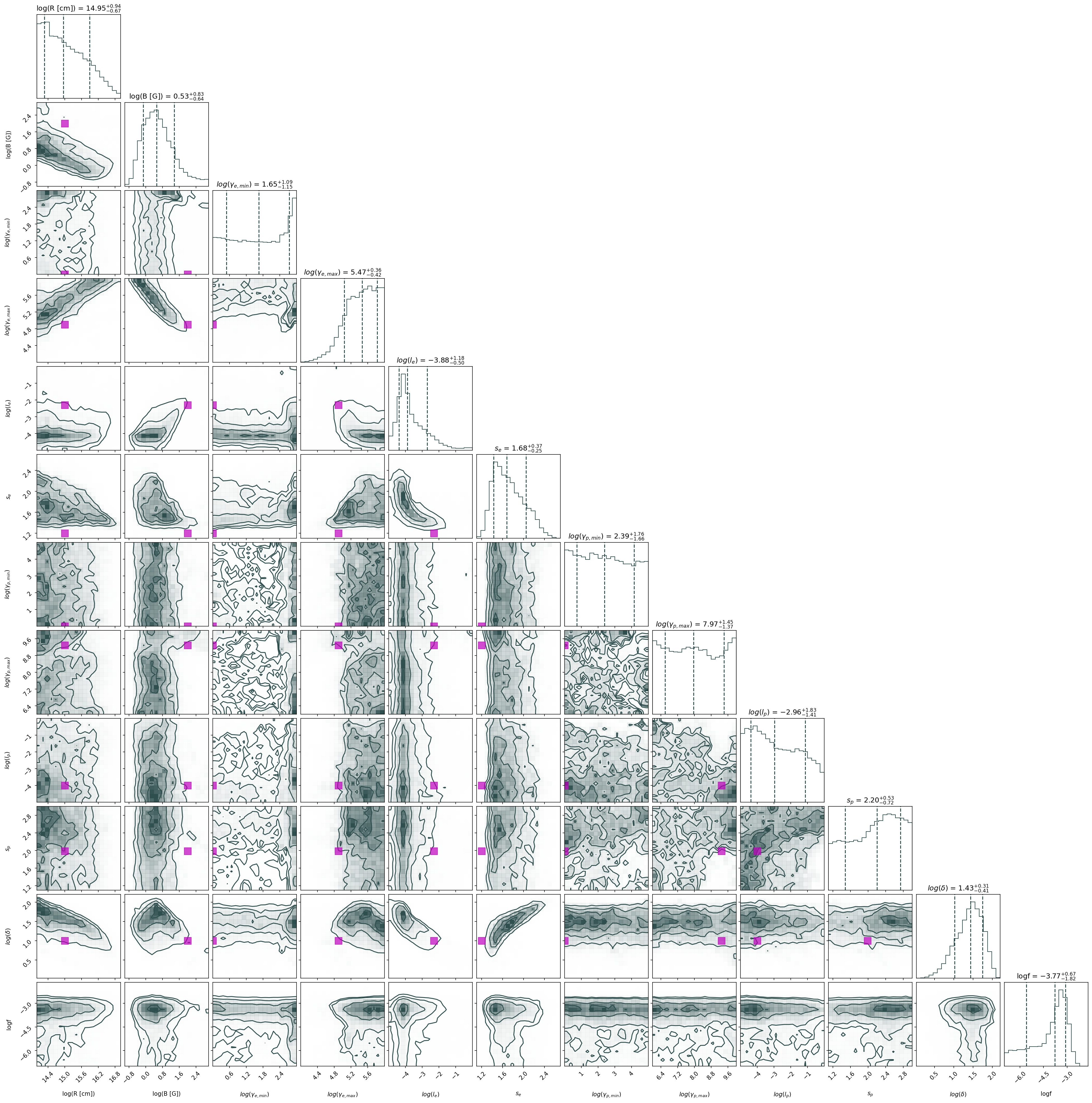} 
\caption{Corner plot showing the posterior distributions of the parameters of the PS model of \hsp. Dashed lines in the histograms indicate the median value of each parameter and the 68 \% range of values. The parameter values used in the PS from \citet{Petro2020} are overplotted (magenta square) for comparison.}
\label{fig:corner-psyn}
\end{figure*}

We present corner plots showing the posterior distributions of parameters for an SSC and a proton-synchrotron model for blazar \hsp. Inspection of the corner plot in Fig.~\ref{fig:corner-ssc} shows that most parameters of the SSC model are well-constrained. However, the radius of the source cannot be pinned down by the SED fitting, suggesting that more information is needed. For example, knowledge of the variability timescale could help eliminate further the posterior distribution. To better understand why the posterior distribution of the radius is not very peaked, let us consider the following. Assuming that electrons are radiating away via synchrotron radiation their injected luminosity, then the synchrotron luminosity in the observer's frame would scale as $L_{syn} \propto \delta^4 L^{inj}_{e} \propto \delta^4 \ell_e/R$. Given that the 68 \% of values for $\delta$ and $\ell_e$ span a range of about $10^{0.6}$ and $10$ respectively, the radius should range within a factor of $\sim 10^{3.4}$ for a fixed observed luminosity. This range is roughly equal to the one used in the prior distribution of $R$, thus explaining the broad range of radii that yield acceptable fits to the data. Regarding the minimum electron Lorentz factor, its posterior distribution is bimodal, with most solutions favoring higher values.  Stronger constraints on this parameter could be placed if observations at GHz frequencies and soft $\gamma$-rays ($10^{21}-10^{22}$~Hz) would be available, as these ranges probe the lower energy radiating electrons. When a model with more parameters is applied to the same data, larger uncertainty in the derived parameter values is expected. This is demonstrated in the corner plot for the PS model (Fig.~\ref{fig:corner-psyn}), where one can see that the posterior distributions for the parameters describing the proton distribution (e.g. $\gamma_{p, \min}$, $\gamma_{p, \max}$, $s_p$)  are very similar to their priors. Some solutions could be excluded a posteriori by implying physical constraints (e.g. by requiring the gyroradius of the maximum energy protons to be smaller than the source size). Nevertheless, it becomes clear that in physical models where the low- and high-energy components of the SED are produced by different particle populations, the parameter space that yields acceptable fits to the data is large. 

\end{document}